\documentclass[a4paper,11pt]{article}
\pdfoutput=1 

\usepackage{jheppub} 

\usepackage[T1]{fontenc} 

\usepackage{amsmath}
\usepackage{amssymb}
\usepackage{graphicx}
\usepackage{bm}
\usepackage{amsfonts}
\usepackage{dsfont}

\usepackage{bbm}
\usepackage{setspace}
\usepackage{physics}
\usepackage{enumitem}
\usepackage[font=small]{caption}
\usepackage{subcaption}

\title{\boldmath On the replica structure of Sachdev-Ye-Kitaev model}

\author[a]{Hanteng Wang,}
\author[b]{D. Bagrets,}
\author[c]{A. L. Chudnovskiy, \footnote{Corresponding author.}}
\author[a,d]{and A. Kamenev}

\affiliation[a]{School of Physics and Astronomy, University of Minnesota, Minneapolis, MN 55455, USA}
\affiliation[b]{Institut f\"ur Theoretische Physik, Universit\"at zu K\"oln, Z\"ulpicher Stra\ss e 77, 50937 K\"oln, Germany}
\affiliation[c]{1. Institut f\"ur Theoretische Physik, Universit\"at Hamburg,
Jungiusstr. 9, D-20355 Hamburg, Germany}
\affiliation[d]{William I. Fine Theoretical Physics Institute, University of Minnesota, Minneapolis, MN 55455, USA}

\emailAdd{wang6243@umn.edu}
\emailAdd{dmitry.bagrets@uni-koeln.de}
\emailAdd{achudnov@physik.uni-hamburg.de}
\emailAdd{kamenev@physics.umn.edu}

\abstract{We investigate existence of  replica off-diagonal solutions in the field-theoretical description of Sachdev-Ye-Kitaev model. To this end we evaluate a set of local and non-local dynamic correlation functions in the long time limit. We argue that the structure of the soft-mode Schwarzian action is qualitatively different in replica-diagonal vs. replica-off-diagonal scenarios, leading to distinct long-time predictions for the correlation functions. We then evaluate the corresponding correlation functions numerically and compare the simulations with analytical predictions of replica-diagonal and replica-off-diagonal calculations. We conclude that all our numerical results are in a quantitative agreement with the theory based on the replica-diagonal saddle point plus Schwarzian and massive Gaussian fluctuations (the latter do contain replica off-diagonal components). This seems to exclude any contributions from replica-off-diagonal saddle points, at least on the time scales shorter than the inverse many-body level spacing.   
}

\begin{document} 
\maketitle
\flushbottom 

\section{Introduction}
\label{sec:intro}

Sachdev-Ye-Kitaev (SYK) model \cite{Sachdev-Ye,kitaev2015talk} has recently attracted a lot of attention as an explicit example of the holographic correspondence \cite{almheiri2015models,cotler2017black,engelsoy2016investigation,gross2017bulk,hartnoll2018holographic,Sachdev,Jensen16,jevicki2016bi,Kitaev}. It also turned out to be a convenient tool to investigate thermalization and chaos 
\cite{maldacena2016bound,you2017sachdev,cotler2017black,bagrets2017power,altland2018quantum,sonner2017eigenstate,krishnan2017quantum,liu2017disorder,eberlein2017quantum,gharibyan2018onset,cotler2017chaos} in the many-body framework. A number of applications towards condensed matter physics    \cite{banerjee2017solvable,song2017strongly,gu2017energy,gu2017local,davison2017thermoelectric,bi2017instability,zhang2018topological,chen2017competition} as well as certain interesting generalizations  \cite{berkooz2017higher,Gross17,fu2017supersymmetric,li2017supersymmetric,witten2016syk,gurau2017complete,klebanov2017uncolored,peng2017supersymmetric,bi2017instability,garcia2018chaotic,nosaka2018thouless} were proposed.    
By now there is a firm understanding of its many-body density of states   \cite{Garcia16,Garcia17,fu2016numerical}, 
level-statistics \cite{you2017sachdev,altland2018quantum}, and certain correlation functions \cite{Polchinski-Rosenhaus2016,Bagrets-Altland-Kamenev2016,bagrets2017power,stanford2017fermionic,Mertens2017,Mertens2018,Lam2018,Blommaert2018}.   

The model is represented by interacting Majorana fermions with quenched random matrix elements. 
As such, it naturally admits a description in terms of a replica field theory \cite{edwards1975theory}. The structure of this theory in the replica space 
has important bearings on all aspects and applications of the SYK-like models. In particular, an existence and properties of a glass phase is most naturally discussed in terms of the replica symmetry breaking (RSB). Indeed, RSB was first introduced by Parisi \cite{parisi1983order,mezard1987spin} to describe glass transition in the Sherrington-Kirkpatrick model \cite{sherrington1975solvable}.  The existence of the glass phase in SYK and similar models is a subject of intense discussions since the very introduction of the model  
\cite{Sachdev-Ye,Georges01,fu2016numerical,ye2018two,qi2018schwarzian}.  A recent discussion, 
Ref.~\cite{gur2018does},  came on the side of the absence of the glass phase. 

Non-trivial replica structures, associated with some form of RSB, were discussed in many other fields of the statistical physics. Most relevant to the present context are replica studies of the random matrix theory. The latter may be classified as SYK$_2$ model (as opposed to SYK$_4$, discussed in this paper). The corresponding replica filed theory of SYK$_2$ is known as non-linear sigma model. Its long time (i.e. small energy) correlation functions were understood in terms of the broken replica symmetry \cite{kamenev1999wigner,kamenev1999level,yurkevich1999nonperturbative,nishigaki2002replica,kanzieper2002replica}. To some extent these studies mirror Altshuler-Andreev description \cite{andreev1995spectral} in terms of the broken supersymmetry. The important point is that RSB is only noticeable on the time scale associated with the inverse level spacing (Heisenberg time) and practically does not have consequences at shorter times.         

It was recently suggested \cite{Stanford_2replicas}  that non-trivial replica structure (i.e. replica non-diagonal saddle point) may be responsible for the behavior of the structure factor of SYK$_4$ model  at a time scale parametrically shorter than the Heisenberg time. Another recent study \cite{Khramtsov} discusses thermodynamic relevance of the replica off-diagonal saddle points in SYK model. 

The goal of this paper is to investigate possible signatures of RSB and replica off-diagonal saddles on the behavior of correlation functions at moderately long times. By those we understand time scales longer than $\tau > N/J$, yet shorter 
than the Heisenberg time (i.e. inverse many-body level spacing).  The correlation functions considered here are motivated by mesoscopic physics \cite{mesoscopic}, where one is interested in quench disorder averages of higher moments of certain quantum observables. (One may also look for an entire probability distribution function of a given 
observable over quench disorder realizations.) Here we show that the corresponding correlation functions exhibit qualitatively distinct behavior being calculated on replica diagonal vs. replica off-diagonal saddle points. The difference stems from the ways the corresponding saddle points break the reparameterization symmetry \cite{kitaev2015talk,CommentsSYK16}  of the model. The distinct patterns of the symmetry breaking are reflected in the structure of the low-energy (Schwarzian) action. We found that in case of replica-diagonal saddle points the latter consists of $n$ (number of replica) independent Schwarzians. However, for a generic replica off-diagonal saddle point there is only one Schwarzian degree of freedom, while the remaining $n-1$ acquire a stiffer action. These observations translate into a different behavior of correlation functions at moderately long times. 

We then perform a detailed comparison of our analytical expectations with  numerical simulations of $N\leq 32$  SYK$_4$ model. Our simulations use exact diagonalization and exact matrix elements to evaluate corresponding ``mesoscopic'' correlation functions. We consider $p$-th moments, $p=1,\ldots, 5$, of both site-local and site-non-local two-point correlation functions.  The comparison shows no evidence for contributions from replica off-diagonal saddles. On the contrary, all the data may be quantitatively accounted for by the theory based on replica diagonal saddle point along with reparameterization fluctuations and massive Gaussian fluctuations around it. The massive fluctuations, which include replica off-diagonal components, must be retained to account for small site non-local correlations.       

This work is organized as follows: in section \ref{sec:model} we review SYK model and its  description in terms of the replica field theory. In section \ref{sec:reparameterizations} we discuss soft reparameterization modes and how their action is different between replica diagonal and replica off-diagonal saddle point configurations. In section \ref{sec:site-local} we discuss consequences of these differences for the long-time behavior of mesoscopic correlation functions. We put these differences to numerical test in section \ref{secSite-local}. In section \ref{sec_Site-non-local} the similar program is implemented to a different family of mesoscopic correlation functions - those with site non-local correlations. We present brief conclusions in section \ref{sec:conclusions}.  A number of technical details are relegated to appendices A-F.  

\section{Model and definitions}
\label{sec:model}

The real-fermion version of SYK$_4$ model  is formulated in terms of Majorana fermions $\chi_i$ on lattice sites $i$. It is determined by the Hamiltonian  
\begin{equation}
\hat{H}=\sum_{i,j,k,l}^N J_{ijkl} \chi_i\chi_j\chi_k\chi_l, 
\label{SYK}
\end{equation}
where matrix elements $J_{ijkl}$ are random statistically independent Gaussian distributed variables with zero mean and a variance given by 
\begin{equation}
\langle (J_{ijkl})^2\rangle =3! J^2/N^3. 
\label{J_ijkl}
\end{equation}
The standard field theoretical treatment of the SYK model \cite{Bagrets-Altland-Kamenev2016, CommentsSYK16} employs the replica-trick. It replicates the fermionic degrees  of freedom as $\chi_i\to \chi_i^a$, where $a=1,\ldots n$, allowing for a direct averaging over the random couplings $J_{ijkl}$.  To arrive at an effective bosonic field theory, describing the  behavior of the model at low energies and long times, one then integrates the fermionic degrees of freedom,  by introducing a replica-matrix valued field   
\begin{equation} 
G^{ab}_{\tau \tau'}=-\frac{1}{N}\sum_{i=1}^N \chi^a_i(\tau)\chi^b_i(\tau'). 
\label{def_Gab}
\end{equation}
Eq. (\ref{def_Gab}) is enforced  by inserting the functional $\delta$-function in the replicated partition function 
\begin{equation}
1=\int[D\Sigma^{ab}_{\tau\tau'}][DG^{ab}_{\tau\tau'}]e^{N \Sigma^{ab}_{\tau\tau'}\left(G^{ab}_{\tau, \tau'}+\sum_{i=1}^N \chi^a_i(\tau)\chi^b_i(\tau')\right)},  
\label{delta-function}
\end{equation} 
where  the matrix field $\Sigma^{ab}_{\tau\tau'}$ is the Lagrangian multiplier dual to $G^{ab}_{\tau \tau'}$.
After integration of the fermionic degrees of freedom, one arrives at the following action 
\begin{equation}
-S[\Sigma, G]=\frac{N}{2}\left[\mathrm{Tr}\ln\left(\partial_{\tau}\delta^{ab}+\Sigma^{ab}_{\tau\tau'}\right)+ 
\frac{J^2}{4}\left[G^{ab}_{\tau\tau'}\right]^4+\Sigma^{ba}_{\tau'\tau}G^{ab}_{\tau\tau'}\right]. 
\label{SYC-Action}
\end{equation}
In the large $N$ limit, where $N$ is the number of sites, the properties of the model are determined by the saddle point of the path integral over the effective fields $G^{ab}_{\tau \tau'}$ and  $\Sigma^{ab}_{\tau\tau'}$. The corresponding saddle point equations read 
\begin{eqnarray}
\Sigma^{ba}_{\tau'\tau} = - J^2 (G^{ab}_{\tau\tau'})^3 \label{Sigma_G3}, \quad \quad 
\big( {\bf\hat{1}} \partial_\tau + \hat {\bf \Sigma}\big) \circ\hat {\bf G}=- {\bf\hat{1}}, 
								\label{EqSigmaG}
\end{eqnarray}
where ${\bf\hat{1}}=\delta_{ab}\delta(\tau-\tau')$. At this junction the standard  choice \cite{Polchinski-Rosenhaus2016,Bagrets-Altland-Kamenev2016,engelsoy2016investigation,stanford2017fermionic,Mertens2018} is to look for a replica-diagonal saddle point solution of the form  $\hat {\bf G} = \delta_{ab} G_{\tau\tau'}$ and correspondingly $ \hat {\bf \Sigma}= \delta_{ab} \Sigma_{\tau\tau'}$. Let us emphasize that, although we call such a choice {\em replica-diagonal}, fluctuations around the replica-diagonal saddle may and should include replica-off-diagonal components $\delta G^{ab}_{\tau\tau'}$. We discuss them in detail in Appendix B. 

However, one may look for genuinely replica-off-diagonal solutions of Eqs.~(\ref{EqSigmaG}). In this paper we restrict ourselves to separable solutions, where matrix form in replica and time spaces separates as:
  \begin{eqnarray}
\hat {\bf G} = g^{ab}G_{\tau'\tau},    \quad\quad 	 \hat {\bf \Sigma} = \sigma^{ab}\Sigma_{\tau'\tau}.							\label{EqSigmaG-separate}
\end{eqnarray}
Here $G_{\tau'\tau}$ and $\Sigma_{\tau'\tau}$ are traditional replica-diagonal solutions, while time-independent {\em symmetric} $n\times n$  
matrices ${\bf g}$ and $\boldsymbol{\sigma} $ satisfy: 
\begin{eqnarray}
\sigma_{ab}= \big(g_{ab}\big) ^3, \quad\quad 
\boldsymbol{\sigma} \cdot {\bf g}={\bf 1}.
										\label{MatrixSP}
\end{eqnarray}
This particular form is motivated by the fact that it allows to naturally keep the conformal structure of the long-time 
effective theory \cite{kitaev2015talk} - the feature that was  proven to be central to the physics of the SYK model. Specifically, 
in the long-time limit, one may neglect the $\partial_\tau$ term   in the saddle point equation Eq. (\ref{EqSigmaG}) and 
find a conformal solution of the form: 
\begin{eqnarray}
\hat {\bf G}=-{\bf g}\,  \frac{J^{-1/2}}{(4\pi)^{1/4}}\frac{\mathrm{sgn}(\tau-\tau')}{|\tau-\tau'|^{1/2}}, 
\quad\quad 
\hat {\bf \Sigma} =-\boldsymbol{\sigma}\,  \frac{J^{1/2}}{(4\pi)^{3/4}}\frac{\mathrm{sgn}(\tau-\tau')}{|\tau-\tau'|^{3/2}},  														\label{ansatz_Sigma} 
\end{eqnarray}
where replica matrices ${\bf g}$ and $\boldsymbol{\sigma} $ satisfy Eqs.~(\ref{MatrixSP}). 

The $n\times n$, where $n$ is number of replica, matrix equations ~(\ref{MatrixSP}) admit a wealth of both diagonal and off-diagonal solutions. It is thus necessary to spell out selection criteria on which of these solutions should be taken into account and why. The most natural of such criteria seems to be a requirement of having a minimal action (i.e. free energy). 
In particular one may ask if the widely accepted choice ${\bf g}= \boldsymbol{\sigma}=\delta_{ab} $ indeed has  the smallest action.   In Appendix A we show that one can find a discrete set of solutions of the form 
\begin{equation}
					\label{eq:RSBansatz}
g_{ab}=\tilde{g}\delta_{ab}+g(1-\delta_{ab}), 					
\end{equation}
where $\tilde g$ and $g$ are $n$-dependent complex numbers. Moreover, in the $n\to 0$ limit the (real part of)  corresponding action is {\em smaller} than that on  the diagonal (i.e. $\tilde g=1$, $g=0$) solution. Similar conclusions were recently reached in Ref.~\cite{Khramtsov}. The question thus arises whether these (or others) replica-off-diagonal solutions are indeed relevant for the physics of the model.   

This question is farther complicated by the fact that besides the saddle point action one needs to evaluate fluctuation determinants and perform summation over the set replica-off-diagonal saddles for any desired observable.  Since we do not know how to perform this program in general, we seek for generic signatures, which help to distinguish between diagonal and off-diagonal solutions. Below we argue that long time behavior of certain correlation functions serves as a sensitive test for the presence of the off-diagonal components. To argue why this is indeed the case one needs to consider a structure of soft-mode fluctuations around diagonal and off-diagonal solutions. For the conformal solutions of the form Eq.~(\ref{ansatz_Sigma}) such soft modes are given by reparameterization  fluctuations \cite{kitaev2015talk,Polchinski-Rosenhaus2016,Bagrets-Altland-Kamenev2016,engelsoy2016investigation,stanford2017fermionic,Mertens2018}.     
  
\section{Reparameterization  fluctuations}
\label{sec:reparameterizations}

In the conformal limit (i.e. neglecting $\partial_\tau$ term) the action (\ref{SYC-Action}) and the saddle point equations Eqs. (\ref{Sigma_G3}),  are invariant under the time reparametrization transformations 
\begin{eqnarray}
\label{rep-transform}
&&G^{ab}(\tau_1,\tau_2)\rightarrow [f'_a(\tau_1)]^{1/4} G^{ab}(f_a(\tau_1), f_b(\tau_2)) [f'_b(\tau_2) ]^{1/4},\\
&&\Sigma^{ab}(\tau_1,\tau_2)\rightarrow [f'_a(\tau_1)]^{3/4} \Sigma^{ab}(f_a(\tau_1), f_b(\tau_2)) [f'_b(\tau_2) ]^{3/4},
\end{eqnarray} 
where $G^{ab}(\tau_1,\tau_2)$ and $\Sigma^{ab}(\tau_1,\tau_2)$ are conformal solutions (\ref{ansatz_Sigma}). 
Here $f_a(\tau)$ with $a=1,\ldots n$ is a {\em replica-specific}  reparametrization transformation. 
This defines the symmetry group ${\cal G}$ of the action~(\ref{SYC-Action}) in the infra-red limit, 
${\cal G} = \otimes_{a=1}^n {\rm Diff}(\mathds{R})$, where ${\rm Diff}(\mathds{R})$ denotes the diffeomorphism group of
time axis. The product over replicas reflects the fact that reparametrization transformations 
can be chosen independently in different replicas, i.e. $f_a(\tau)\neq f_b(\tau)$. 
The symmetry under time-reparametrizations is a crucial property, that relates the SYK model to the AdS$_2$ gravity theories \cite{almheiri2015models,Maldacena16,cotler2017black,engelsoy2016investigation,gross2017bulk,hartnoll2018holographic,Sachdev,Jensen16,jevicki2016bi,Kitaev}. This time-reparametrization symmetry is however spontaneously broken
by the saddle point solutions~(\ref{ansatz_Sigma}) down to the subgroup 
$H={\rm SL}(2,\mathds{R}) \subset {\cal G}$  
resulting in the appearance of a soft modes,  which
span the coset ${\cal G}/H$.  Specifically, the group $H$ is formed by all M\"obius maps of the form 
$h(\tau)=(A \tau + B )/(C\tau + D) $ with $AD - BC = 1$. 

Here, the major difference shows up between the diagonal and off-diagonal saddle point solutions.  
For the diagonal case the subgroup is $\widetilde H = \otimes_{a=1}^n{\rm  SL}(2, {\mathds R})$. Indeed,
for a diagonal saddle point the independent  M\"obius maps  $h_a(\tau)=(A_a\tau + B_a )/(C_a\tau + D_a)$ may be taken for each replica, leaving the diagonal solution  (\ref{ansatz_Sigma}) invariant. The diagonal soft mode coset is 
thus ${\cal G}/\widetilde H=\big[\otimes_{a=1}^n  {\rm Diff}(\mathds{R})\big] /\big[\otimes_{a=1}^n{\rm  SL}(2, {\mathds R})\big]$. 

This should be contrasted with the off-diagonal case, where the subgroup is  $H = {\rm  SL}(2, {\mathds R})$ - the {\em same} for all replicas. Indeed, performing different M\"obius transformations in different replicas does {\em not} leave 
(\ref{ansatz_Sigma})  invariant, if ${\bf g}$ and $\boldsymbol{\sigma}$  have off-diagonal components\footnote{We are grateful to Mikhail Khramtsov for discussing this point.}. Therefore the coset is different: ${\cal G}/ H=\big[\otimes_{a=1}^n  {\rm Diff}(\mathds{R})\big]/{\rm  SL}(2, {\mathds R})$. The different structure of the coset is reflected in the soft mode action.

The latter action originates from the explicit breaking of the reparametrization symmetry by the time derivative term 
$ \delta_{ab} \partial_\tau$. In the diagonal case, where the coset is the  product of $n$ independent  ${\rm Diff}(\mathds{R})/ {\rm  SL}(2, {\mathds R})$ components, the corresponding action is the sum of $n$   
Schwarzian derivatives ~\cite{CommentsSYK16} 
\begin{equation}
							\label{eq:Sdiag}
S_\mathrm{diag}=-M\sum\limits_{a=1}^n \int \!d\tau\,  \mathrm{Sch}(f_a,\tau)
\end{equation}
where $M\sim N \log N$ is the mass of the soft fluctuations \cite{Bagrets-Altland-Kamenev2016}. 

In the off-diagonal case the subgroup $H$ consists of a single ${\rm  SL}(2, {\mathds R})$, suggesting that only a single 
degree of freedom is governed by the Schwarzian action.  
Indeed, the explicit calculation, outlined  in details in  Appendix E, shows that  the off-diagonal matrix elements $g_{ab}\neq 0$ in the saddle point solution generate additional terms in the action for reparametrization fluctuations. These terms  overpower $n-1$ Schwarzian derivatives in the long-time limit. 

Let us use the exponential representation of reparametrizations \cite{Bagrets-Altland-Kamenev2016}
\begin{equation}
f_a(\tau)=\int^{\tau} \exp[ \phi_a(\tau)] d\tau,
						\label{expRep}
\end{equation} 
which has an advantage that the corresponding invariant integration measure is flat in $\phi_a$ variables.   
In this representation,  the additional action  can be cast in the form of an effective potential: 
\begin{equation}
           \label{eq:S2_line1}
S_2[\phi]=- \frac{N J }{2^7 \sqrt{2 \pi}} \sum_{a\neq b} g^2_{ab}\int\limits_{\cal C}  
\frac{ d\tau }{ \cosh^{3/2} \left[\phi_a(\tau_1(\tau))- \phi_b(\tau_2(\tau))\right]}.
\end{equation} 
Here the integration goes along the line ${\cal C}=(\tau_1(\tau), \tau_2(\tau))$ drawn in $\mathds{R}^2$ space of two times, 
at which two reparametrizations take equal values,  $f_a(\tau_1(\tau))=f_b(\tau_2(\tau))$. When expanded in small deviations $\phi_a- \phi_b$,  each term in the action Eq. (\ref{eq:S2_line1}) acquires the form of a ``mass'' term 
\begin{equation}
S_2[\phi]\simeq 
 \frac{5 N J }{2^{10} \sqrt{2 \pi} }  \sum_{a\neq b} g_{ab}^2\int 
(\phi_a- \phi_b)^2 d\tau,
\label{S2_mass}
\end{equation}
where in the last expression the integral already goes along the straight line. It is clear that this term is minimized when 
reparametrizations in all replicas are identical and penalizes deviations from such configuration. To formalize this observation we introduce new variables as $\phi_a=\Phi+\varphi_a$,  where $\sum_{a=1}^n \varphi_a \equiv 0$ and therefore $\Phi= {1\over n} \sum_{a=1}^n \phi_a$. Then the soft mode action for,   e.g., off-diagonal ansatz (\ref{eq:RSBansatz}) takes the form  
\begin{equation}
S_\mathrm{off-diag} =- \tilde g^2 M \int \!d\tau\,  {\mathrm{Sch}}(\Phi,\tau) 
+ 
2n g^2 \frac{5 N J }{2^{10} \sqrt{2 \pi} }  \sum_{a=1}^n \int\! d\tau\, \varphi_{a}^2,
 									\label{repAction-full_phi}
\end{equation}
where we have used that $\sum_{a\neq b}(\phi_a-\phi_b)^2=\sum_{ab}(\varphi_a-\varphi_b)^2 =2n\sum_a\varphi_a^2$, since $\sum_a\varphi_a=0$. In the long time limit the last term here suppresses fluctuations of $n-1$ degrees of freedom $\varphi_a$, leaving the single degree of freedom $\Phi$, to be governed by the Schwarzian action. This effectively locks reparameterization degrees of freedom in different replicas to
\begin{equation}
									\label{eq:locking}
f_a(\tau)=f(\tau)=\int^\tau \exp[\Phi(\tau)] d\tau.  
\end{equation}

Finally, let us mention that the structure of action Eq. (\ref{repAction-full_phi}) is consistent with the coset space ${\cal G}/H$
of the replica off-diagonal SYK action. For the infinitesimal reparametrizations $f_a(\tau)= \tau + \epsilon_a(\tau)$ the phases $\phi_a(\tau) \simeq \epsilon'_a(\tau) $. We see that the action Eq. (\ref{eq:S2_line1}), if written in terms of $\epsilon_a(\tau)$, remains {\em massless} vis-a-vis $n$ degrees of freedom. However only single degree of freedom is 
``super soft'': ${\mathrm{Sch}}(\Phi,\tau) \propto ({\cal E}'')^2$ (where ${\cal E} = \sum\limits_{a=1}^n \epsilon_a$), while remaining $n-1$ acquire stiffer action $\propto (\epsilon_a')^2$.   This is {\em not} the case in the diagonal case where all $n$ modes are super soft $\propto (\epsilon_a'')^2$. 

The locking of reparameterization modes in different replicas, Eq.~(\ref{eq:locking}), for off-diagonal saddle points has 
important consequences for long time behavior of the correlation functions, which we explore in the next section. 

\section{Site-local correlation functions}
\label{sec:site-local}

It is well known, that the reparameterization fluctuations modify the long-time decay of  correlation functions. The simplest example is the two-point site-local function:  
\begin{equation}
G(\tau, 0)=\frac{1}{N}\sum_i^N\langle \chi_i(\tau)\chi_i(0)\rangle.
										\label{Corr_chichi}
\end{equation}
While at short times, $1/J<|\tau|<N/J$, the decay is governed by the conformal mean field behavior $G \sim |\tau|^{-1/2}$, Eq.~(\ref{ansatz_Sigma}), its long-time behavior, $|\tau|>N/J$, is very different: $G \sim |\tau|^{-3/2}$ due to the effect of   the reparametrization fluctuations \cite{Bagrets-Altland-Kamenev2016}. 
Moreover, the $2p$-point correlation functions ($p<N$) of the form 
\begin{equation}
G_{2p}(\tau, 0)=\frac{1}{N^p}\sum_{i_1\ldots i_p}^N  \langle \chi_{i_1}(\tau)\ldots\chi_{i_p}(\tau)\chi_{i_1}(0)\ldots\chi_{i_p}(0) \rangle.
										\label{Corr_chichichichi}
\end{equation}
at long time decay with the {\em same} universal exponent $-3/2$, i.e. $G_{2p} \sim |\tau|^{-3/2}$ \cite{Bagrets-Altland-Kamenev2016}. The short time behavior is, of course, $p$-dependent: $G_{2p} \sim |\tau|^{-p/2}$.  It is important to notice that the angular brackets 
in Eqs.~(\ref{Corr_chichi}) and (\ref{Corr_chichichichi}) imply both quantum mechanical ground-state expectation value  (hereafter we restrict ourselves to zero temperature)  along with the averaging over disorder realizations. 

We now introduce different objects, inspired by mesoscopic fluctuations physics \cite{mesoscopic}
\begin{equation}
\langle [G(\tau, 0)]^p\rangle_{\mathrm{dis}}=\left\langle\left[N^{-1}\sum_{i=1}^N \langle GS|\chi_i(\tau) 
\chi_i(0)|GS \rangle\right]^{p}\right\rangle_{\mathrm{dis}}, 
										\label{G_p}
\end{equation}
where $|GS \rangle$ stays for a disorder specific ground-state of the SYK$_4$ model (the same for all $p$ expectation values), while $\langle \ldots \rangle_\mathrm{dis}$ denotes averaging over realizations of random matrix elements $J_{ijkl}$. In the replica formalism, the correlation function Eq. (\ref{G_p}) can be written as 
\begin{equation}
\langle [G(\tau,0)]^p\rangle_{\mathrm{dis}}=\left(\frac{1}{N}\right)^p\sum_{i_1\dots i_p=1}^N\left\langle \chi_{i_1}^{a_1}(\tau)  \ldots \chi_{i_p}^{a_p}(\tau)\chi_{i_1}^{a_1}(0)\ldots \chi_{i_p}^{a_p}(0) \right\rangle,   
												\label{Gii_p-replicas}
\end{equation}
where angular brackets denote averaging with respect to the replicated action and all the replicas $a_1, ..., a_p$ are different. The leading contribution to the correlation function Eq. (\ref{Gii_p-replicas}) is given by the product of replica-diagonal contractions. Indeed, each contraction of fermions with different replicas enforces the equality of the sites of the contracted fermions, for example $\langle \chi_{i_1}^{a_1}\chi_{i_2}^{a_2}\rangle\propto g_{a_1a_2}\delta_{i_1 i_2}$ thus eliminating one summation over sites. Such contribution is therefore suppressed by the factor $1/N$ (in case of the replica diagonal saddle point, $g_{a_1a_2}=0$ and such contractions originates from Gaussian fluctuations of $\delta G_{a_1a_2}$ and $\delta \Sigma_{a_1a_2}$, bringing additional factors of $1/N$).  The leading contribution from the product of replica diagonal contractions has furthermore to be averaged over the reparametrization fluctuations 
\begin{equation}
\langle [G(\tau,0)]^p\rangle_{\mathrm{dis}}\approx  \int\prod_{a=1}[D\phi_a(\tau)] \prod_{a=1}^p G^{aa}(f_a(\tau), f_a(0)) \,e^{-S[\phi]},   
				\label{Gii_p-leading}
\end{equation}
where $G^{aa}(f_a(\tau), f_a(0)) $ is given by Eqs. (\ref{rep-transform}), (\ref{expRep}). 

In the case of the off-diagonal saddle point,  the reparametrizations are locked, Eq.~(\ref{eq:locking}), and therefore 
the integration in  Eq.~(\ref{Gii_p-leading}) runs over the single field $\Phi$. This  makes Eqs.~(\ref{Corr_chichichichi}) and (\ref{Gii_p-replicas}) essentially equivalent in the long time regime. One thus expects to find $\langle [G(\tau,0)]^p\rangle_{\mathrm{dis}}\\ \sim \tau^{-3/2}$ independent on $p$. On the other hand, in the replica diagonal case   the reparametrizations are not locked, the integration in  Eq.~(\ref{Gii_p-leading}) runs over $p$ independent field and one expects $\langle [G(\tau,0)]^p\rangle_{\mathrm{dis}}\sim \tau^{-3p/2}$ again in the long time regime. For short times 
reparameterizations are not relevant and one expects mean-field $\langle [G(\tau,0)]^p\rangle_{\mathrm{dis}}\sim \tau^{-p/2}$ irrespective of the replica structure. To summarize: 
\begin{equation}
					\label{eq:local}
\left(\langle [G(\tau,0)]^p\rangle_{\mathrm{dis}}\right)^{1/p} \sim	\begin{cases} \tau^{-1/2}, \quad  \tau<N/J\\
\tau^{-3/2}, \quad   \tau>N/J  \quad \quad  \,\, \mbox{replica diagonal}\\
\tau^{-3/2p} , \quad \tau>N/J \quad\quad  \mbox{replica off-diagonal}.	
\end{cases}			
\end{equation}
This can be checked numerically to distinguish between diagonal and off-diagonal scenario.

\section{Numerical results for site-local correlation functions}
\label{secSite-local}

\begin{figure}[ht]
  \begin{tabular}{@{}cccc@{}}
    \includegraphics[width=.5\textwidth]{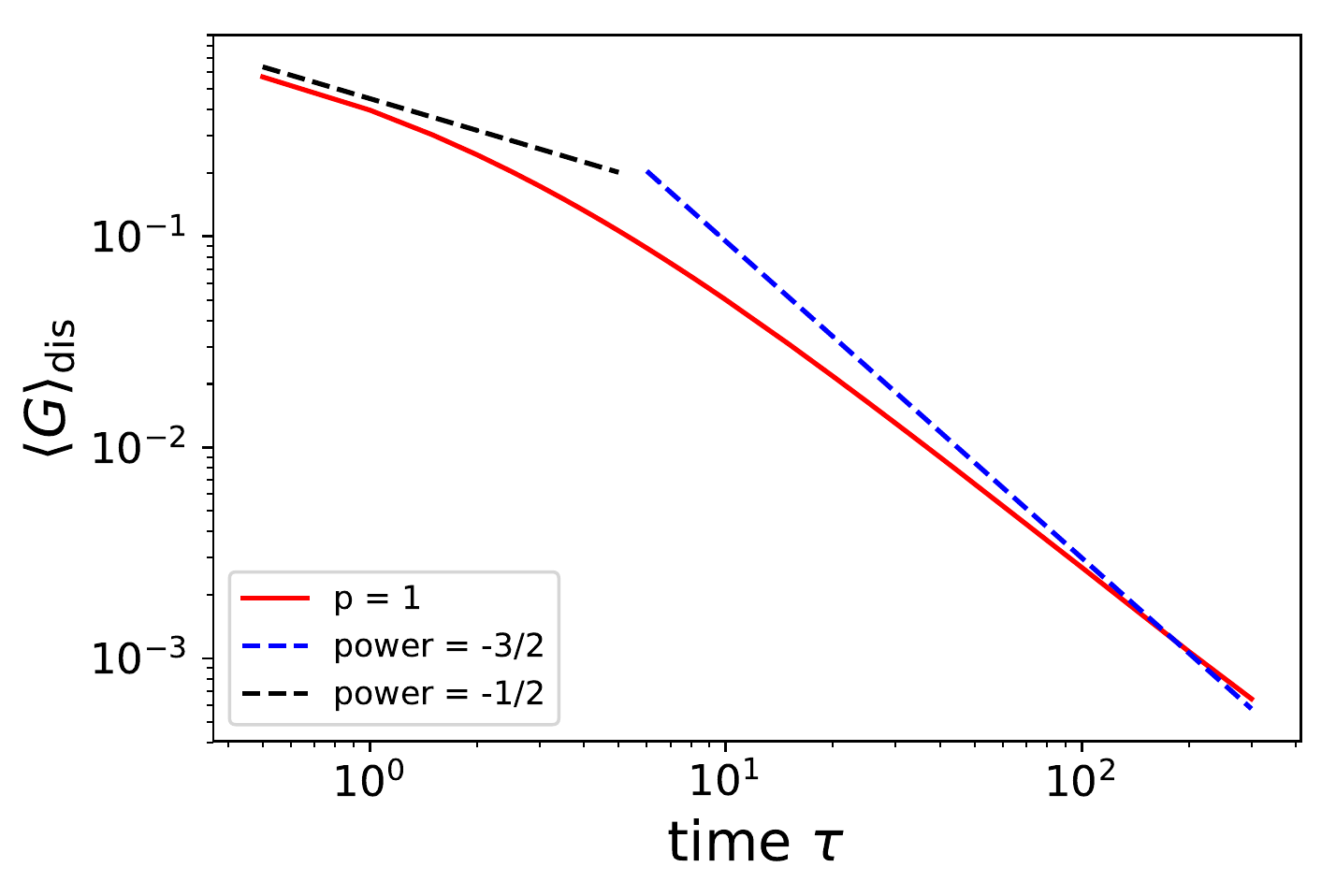} &
    \includegraphics[width=.5\textwidth]{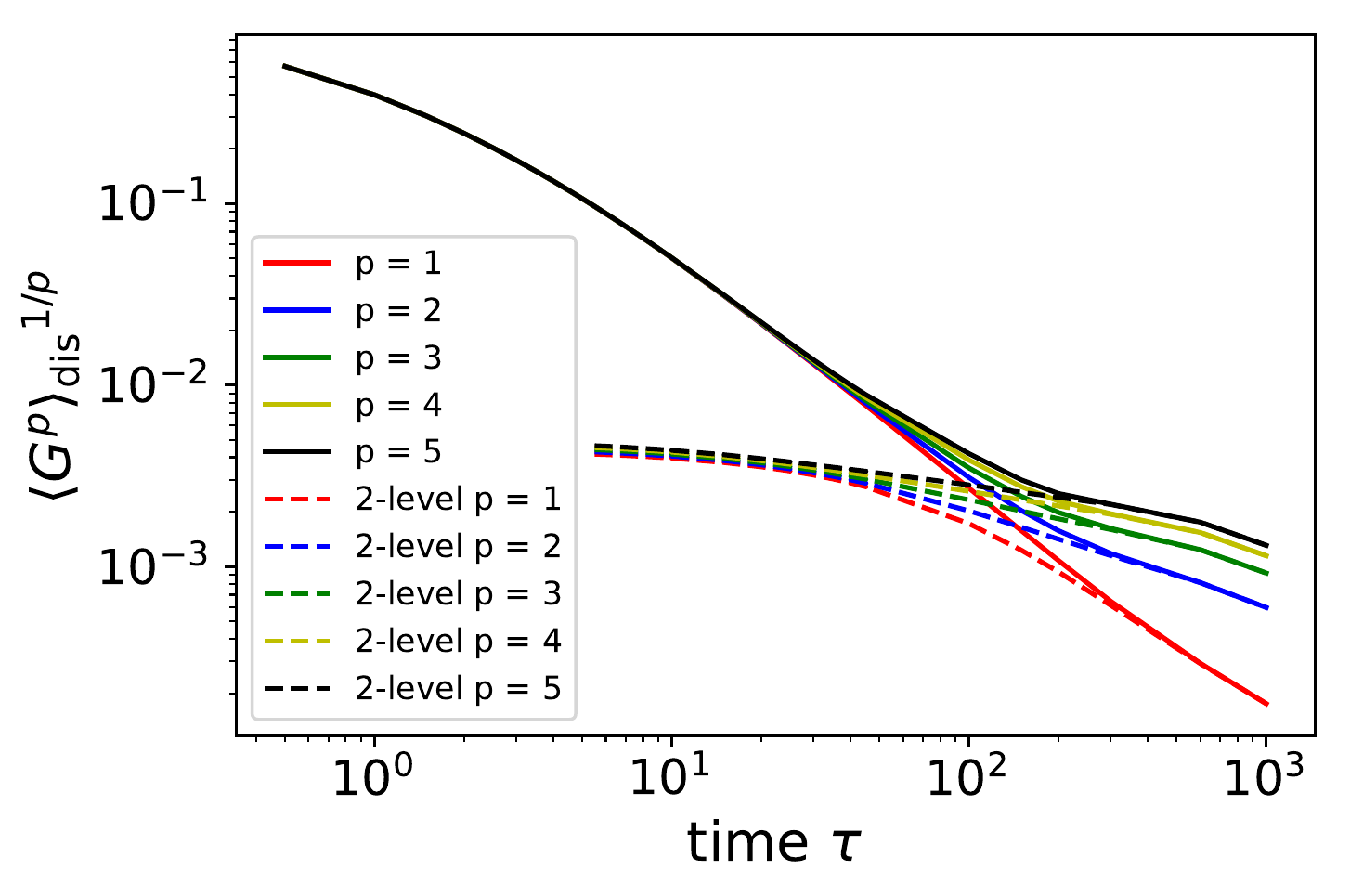}
  \end{tabular}
\caption{(a) Numerical results for $\langle G(\tau,0) \rangle_{\mathrm{dis}}$ at $N=32$ averaged over 30 realizations (Log-Log plot). At short time, it decays with power $-1/2$, while at long time it decays with power $-3/2$. (b) Numerical results for $\langle G(\tau,0)^p \rangle_{\mathrm{dis}} ^{1/p}$ in time domain (Log-Log plot). The dashed lines are Green's functions obtained by taking into account the contribution from the lowest two eigenstates only. Time is measured in units of $1/J$.}
  \label{fig:Gp}
\end{figure}

The basic quantity for numerical calculations is the two-time ground-state expectation value: 
\begin{equation}
G_{ii}(\tau)=\langle GS| \chi_i(\tau)\chi_i(0)|GS\rangle = \sum_n \langle GS|\chi_i|n\rangle\langle n|\chi_i|GS\rangle\,  e^{-(E_n-E_{GS})\tau}. 
\label{defGii}
\end{equation}
In the second equation $|n\rangle$ denote  many-body excited states (with the parity opposite to that of the ground-state).  Numerically, the correlation function Eq. (\ref{defGii}) is calculated from the spectrum of energies and matrix elements obtained by exact diagonalization (see Appendix D for details). 
The correlation function Eq. (\ref{defGii}) is then used to construct the higher order correlation functions as defined by Eq. (\ref{G_p}). 
Numerical results for the correlation function Eq. (\ref{G_p})  
are shown in Fig. \ref{fig:Gp}. 

The correlation function $\langle G(\tau,0)\rangle_{\mathrm{dis}}$ ($p=1$) is shown in the left panel in Fig. \ref{fig:Gp}. Its time-decay exhibits three qualitatively different regimes. At short times ($1\lesssim \tau \lesssim 10$ in units of $1/J$) the correlation function decays as $\tau^{-1/2}$. This behavior corresponds to a saddle point solution, Eq.~(\ref{ansatz_Sigma}). At longer times, for $10\lesssim \tau \lesssim 100$,  the time decay changes to $\tau^{-3/2}$. Such a behavior signals the dominant effect of soft reparametrization fluctuations around the saddle point, as described in Ref. \cite{Bagrets-Altland-Kamenev2016}. At still longer times, $\tau \gtrsim 100$, the time decay of the correlation function is dominated by a first excited many-body state $|n=1\rangle$ (we'll refer to it as "two-level" system), due to the discreteness of the energy spectrum  in a finite size system. The crossover to the two-level regime at long times is quantified  on the right panel of Fig. \ref{fig:Gp}. In that panel, the dashed lines correspond to the calculation of the correlation functions taking into account the two lowest energy levels, $|GS\rangle$ and $|n=1\rangle$, only.  

The right panel in Fig. \ref{fig:Gp} shows that the correlation functions $(\langle [G(\tau,0)]^p\rangle_{\mathrm{dis}})^{1/p}$, calculated for different $p$,  coincide in a wide time range, which includes both mean-field and reparameterization dominated regimes. Comparing this behavior with the theoretical expectations, Eq.~(\ref{eq:local}), we conclude that it is consistent {\em only} with the {\em replica-diagonal} structure of the saddle point. We present an additional independent support to this conclusion by considering site non-local correlation functions in the next section.  

Eventually graphs for different $p$ diverge on approaching the two-level system regime. This latter behavior may be quantitatively explained assuming some (independent) distribution functions for matrix elements $ \langle GS|\chi_i|1\rangle$ and 
energy splitting $E_1-E_{GS}$ (notice that since the ground-state and the excited state belong to different parity sectors, there is no repulsion between them). See appendix F for more details on the two-level regime. To the best of our knowledge, it is not known how to incorporate two-level regime into the replica filed-theory discussed here (see 
Ref.~\cite {altland2018quantum} for an alternative approach).  The situation is very different in SYK$_2$ model, where the corresponding filed-theory  is rotationally invariant in the replica space, allowing for the treatment of RSB at the two-level energy scale \cite{kamenev1999wigner,kamenev1999level,yurkevich1999nonperturbative,nishigaki2002replica,kanzieper2002replica}. 

\section{Site non-local correlation functions}
\label{sec_Site-non-local}

\begin{figure}[ht]
\includegraphics[width=\linewidth]{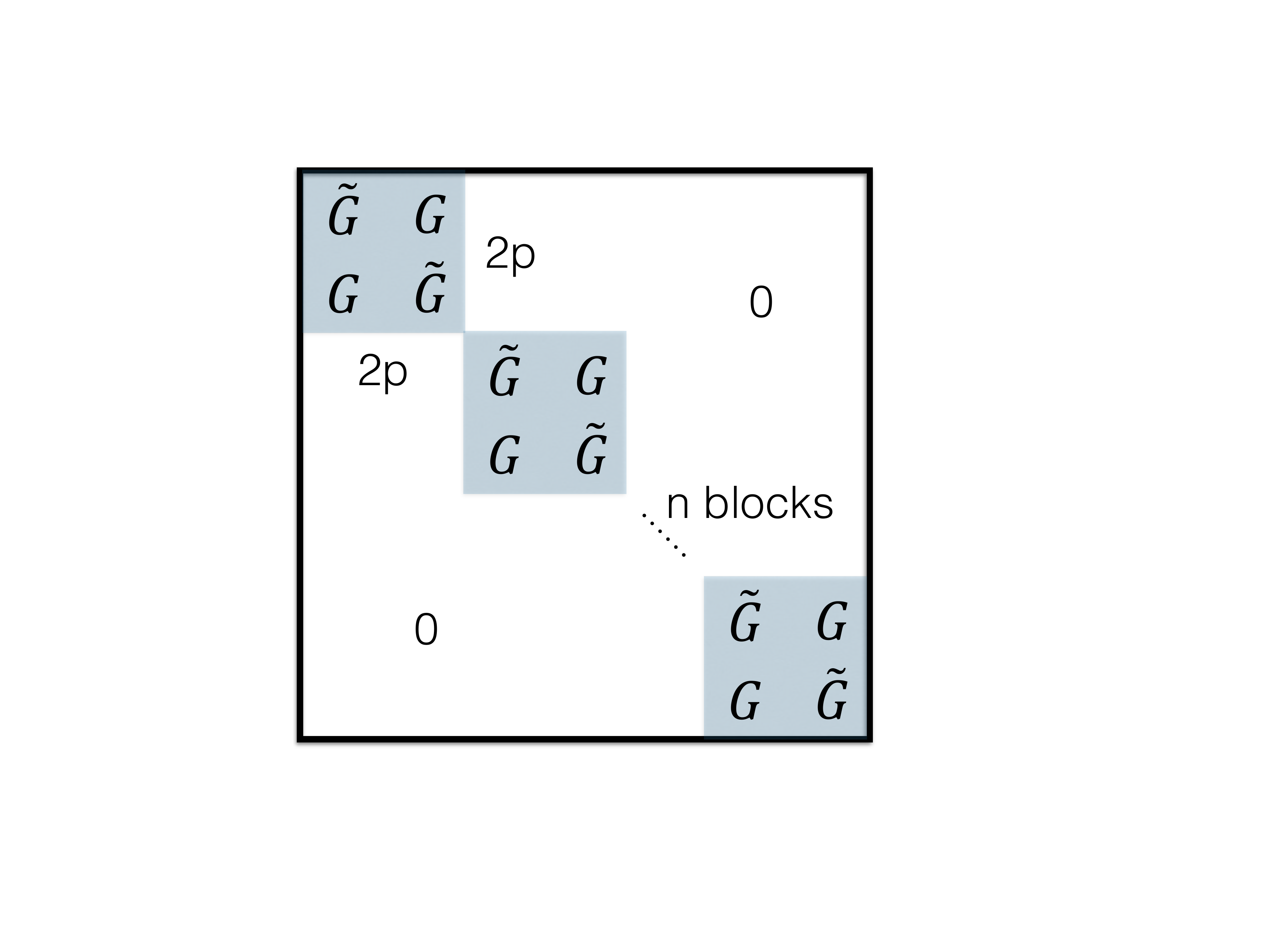}
\vskip -2cm
\caption{Structure of the saddle point matrix for calculation of correlation function $\mathcal{D}_{2p}$}
\centering
\label{RSB-Matrix_2}
\end{figure}

The existence of the replica off-diagonal solutions may be also detected by considering  site non-local  correlation functions of the type: 
\begin{equation}
\mathcal{D}_{2p}(\tau)=\left\langle\left[G_{ij}(\tau,0)G_{ji}(\tau,0)\right]^p\right\rangle_{\mathrm{dis}}=\left\langle\left[\langle GS|\chi_i(\tau)\chi_j(0)|GS \rangle \langle GS| \chi_j(\tau)\chi_i(0)|GS \rangle \right]^p\right\rangle_{\mathrm{dis}},
\label{D2p}
\end{equation}
with $i\neq j$. The advantage of this object is that it vanishes, being calculated at the replica-diagonal saddle point (without account for massive fluctuations), but does {\em not} vanish, being calculated at the off-diagonal saddle point.
To see this we rewrite it  in the replica formalism as, 
\begin{eqnarray}
\nonumber && 
\mathcal{D}_{2p}(\tau)=\left\langle\chi^{a_1}_i(\tau)\chi^{a_1}_j(0)\chi^{a_2}_j(\tau)\chi^{a_2}_i(0)... 
\chi^{a_{2p-1}}_i(\tau)\chi^{a_{2p-1}}_j(0)\chi^{a_{2p}}_j(\tau)\chi^{a_{2p}}_i(0)\right\rangle \\ 
&& 
\approx \Big\langle\chi^{a_1}_i(\tau)\chi^{a_2}_i(0)... 
\chi^{a_{2p-1}}_i(\tau)\chi^{a_{2p}}_i(0)\Big\rangle
\Big\langle\chi^{a_1}_j(0)\chi^{a_2}_j(\tau)... \chi^{a_{2p-1}}_j(0)\chi^{a_{2p}}_j(\tau)\Big\rangle,
\label{D2p_replica}
\end{eqnarray}
where in the second line we disregarded Gaussian fluctuations and used the site-locality of the saddle point correlation functions (both replica diagonal and off-diagonal ones).  Since all replica indexes $a_1,\ldots, a_{2p}$ are distinct here,
it is clear that the second line in Eq.~(\ref{D2p_replica}) is zero on the diagonal saddle point. To estimate it in the replica non-diagonal saddle point we consider block-diagonal matrices ${\bf g}$ and $\boldsymbol{\sigma}$, consisting of $n$ blocks each of the size $2p\times 2p$ along the main diagonal (see Fig. \ref{RSB-Matrix_2}). 
Here we perform explicit calculations for the case $p=1$. Using the saddle point ansatz Eq. (\ref{ansatz_Sigma}), we obtain the correlation function ${\mathcal D}_2$ in the form 
\begin{equation}
\mathcal{D}_{2}(\tau)=\left\langle\chi^{a_1}_i(\tau)\chi^{a_2}_i(0)\right\rangle
\langle\chi^{a_1}_j(0)\chi^{a_2}_j(\tau)\rangle = -\frac{g_{a_1a_2}^2 }{(4\pi)^{1/2}}\frac{1}{|\tau|}, 
\label{D2_SP}
\end{equation}
where we use the saddle point matrix ${\bf g}$ consisting of $2\times 2$ blocks. Replica non-diagonal solutions of Eqs. (\ref{MatrixSP}) for $2\times 2$ blocks read 
\begin{equation}
{\bf g}={\bf 1}_n\otimes \frac{1}{\sqrt{J}2^{1/4}}\left(\begin{array}{cc}
1 & \pm i \\ 
\pm i & 1 
\end{array}
\right) 
, \, \, \, 
\boldsymbol{\sigma}={\bf 1}_n\otimes \frac{1}{2^{3/4}}\left(\begin{array}{cc}
1 & \mp i \\ 
\mp i & 1 
\end{array}
\right) .
\label{22MatrixSP}
\end{equation}
Eq. (\ref{D2_SP}) describes the behavior of the correlation function at short times, when the influence of reparameterization fluctuations is negligible. To obtain the correct time dependence at longer times,  replica off-diagonal correlation function has to be averaged over the reparametrization fluctuations (see Appendix C for details). Since the reparameterizations  are locked according to Eq.~(\ref{eq:locking}), one integrates over a single reparameterization degree of freedom for the both replicas involved in $\mathcal{D}_{2}(\tau)$.  This leads to:  
\begin{equation}
D^{\mathrm{off-diag}}_2(\tau)=\frac{1}{2\sqrt{2\pi}J |\tau|^{3/2}}.  
\label{D2_ROD}
\end{equation} 

We come back now to the replica diagonal scenario, which leads to vanishing result for the non-local functions (\ref{D2p}), being calculated at the diagonal saddle point. However one can include massive (with the mass of order $N$) Gaussian  fluctuations $\delta G_{ab}$ and $\delta \Sigma_{ab}$ around the diagonal saddle point to find a non-zero 
result for the first line in Eq.~(\ref{D2p_replica}) (see Appendix B for detailed derivation). This leads to:  
\begin{equation}
D^{\mathrm{diag}}_2(\tau)= \frac{3}{4\pi 2^{10} J N^3}\frac{1}{|\tau|}
\label{D2_RD}
\end{equation} 
for $\tau< N/J$. 
Subsequent  averaging over the reparameterization fluctuations around the replica-diagonal saddle point with two independent (unlocked) reparameterization modes, one for each replica, results in: 
\begin{equation}
D^{\mathrm{diag}}_2(\tau)= \frac{3}{4\pi 2^{10} J N^3}\frac{1}{|\tau|^{3}}.
\label{D2_RD_rep}
\end{equation}
Analytical calculations of the correlation function Eq. (\ref{D2p_replica}) for $p>2$ result in the following general relation between the correlation functions for different powers $p$ 
\begin{eqnarray}
\nonumber  &&
\left(\frac{\mathcal{D}_{2p}(\tau)}{(2p-1)!!}\right)^{1/p}=\left\{ \begin{array}{ll}
\frac{1}{2\sqrt{2\pi}J |\tau|}, \, \, & \tau<N/J \quad \mbox{replica off-diagonal }, \\ 
\frac{3}{4\pi 2^{10} J N^3}\frac{1}{|\tau|}, \, \, & \tau<N/J  \quad \mbox{replica diagonal}.
\end{array}
\right.   \\ 
\nonumber   && 
\left(\frac{\mathcal{D}_{2p}(\tau)}{(2p-1)!!}\right)^{1/p}=
\left\{ \begin{array}{ll}
\frac{1}{2\sqrt{2\pi}J |\tau|^{3/2p}}, \, \, & \tau>N/J  \quad \mbox{replica off-diagonal}, \\ 
\frac{3}{4\pi 2^{10} J N^3}\frac{1}{|\tau|^{3}}, \, \, &\tau>N/J  \quad  \mbox{replica diagonal }.
\end{array}
\right.   \\
\label{D2p_D2}
\end{eqnarray}
As one can see from Eqs. (\ref{D2_SP}), (\ref{D2_RD}), for short times the two above mentioned scenarios differ  only in the scaling of the correlation function $\mathcal{D}_2(\tau)$ with the number of sites $N$, while for long times both the scaling with $N$ as well as the predicted time dependence become different.  Therefore, the time dependence {\em as well as the dependence on the total number of sites $N$} can be used do discriminate between Eqs. (\ref{D2_ROD}) and  (\ref{D2_RD}). 

\begin{figure}
\includegraphics[width=\linewidth]{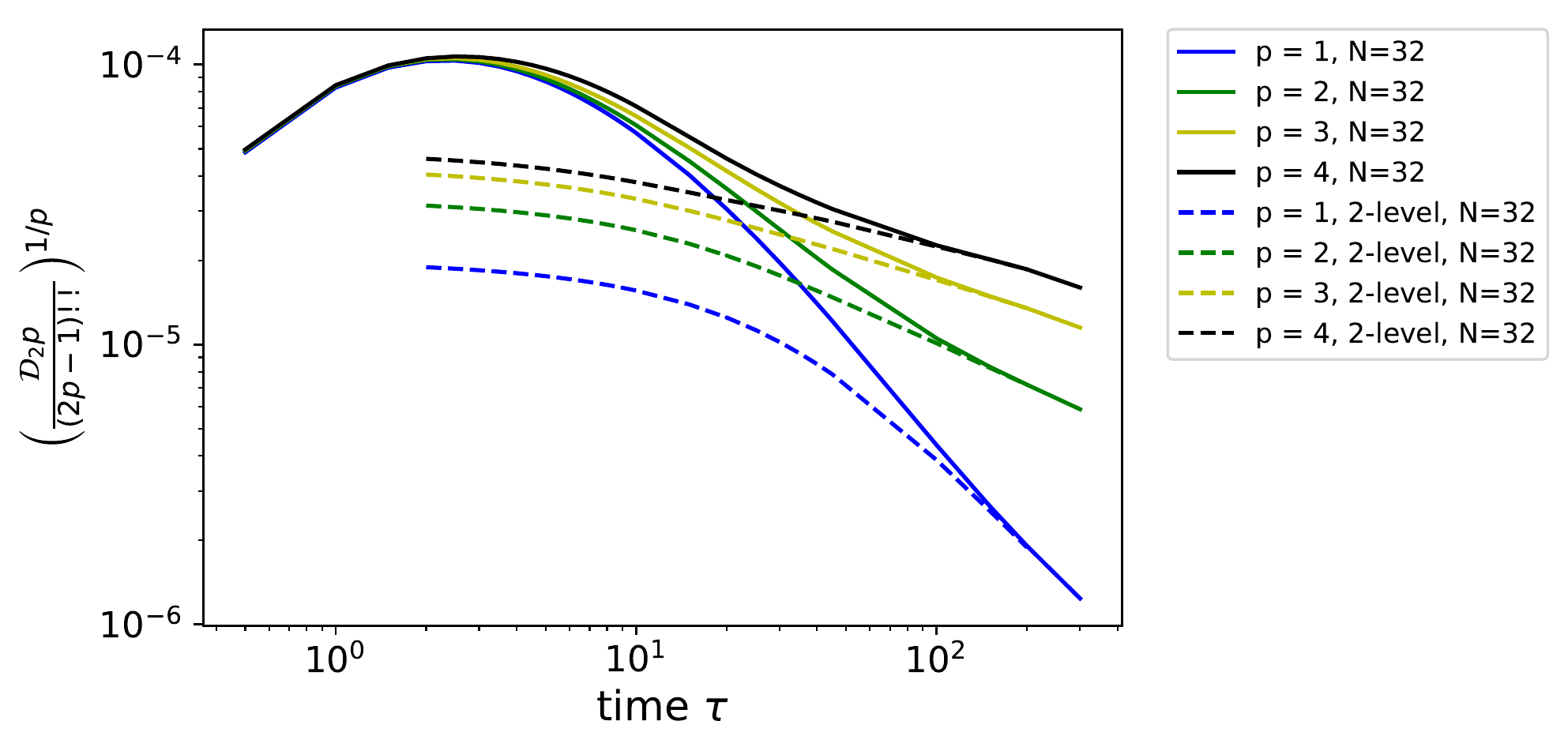}
\caption{Log-Log plot of $\mathcal{D}_{2p}$ versus $\tau$. The $p=1$ graph is consistent with Eqs. (\ref{D2_RD}), (\ref{D2_RD_rep}) between $5\lesssim \tau\lesssim 100$.  For $p\geq 2$ crossover to the two-level regime is too fast to deduce the time dependence expected from Eq.~(\ref{D2p_D2}). }
\label{fig_D2p-tau}
\end{figure}
\begin{figure}
\includegraphics[width=\linewidth]{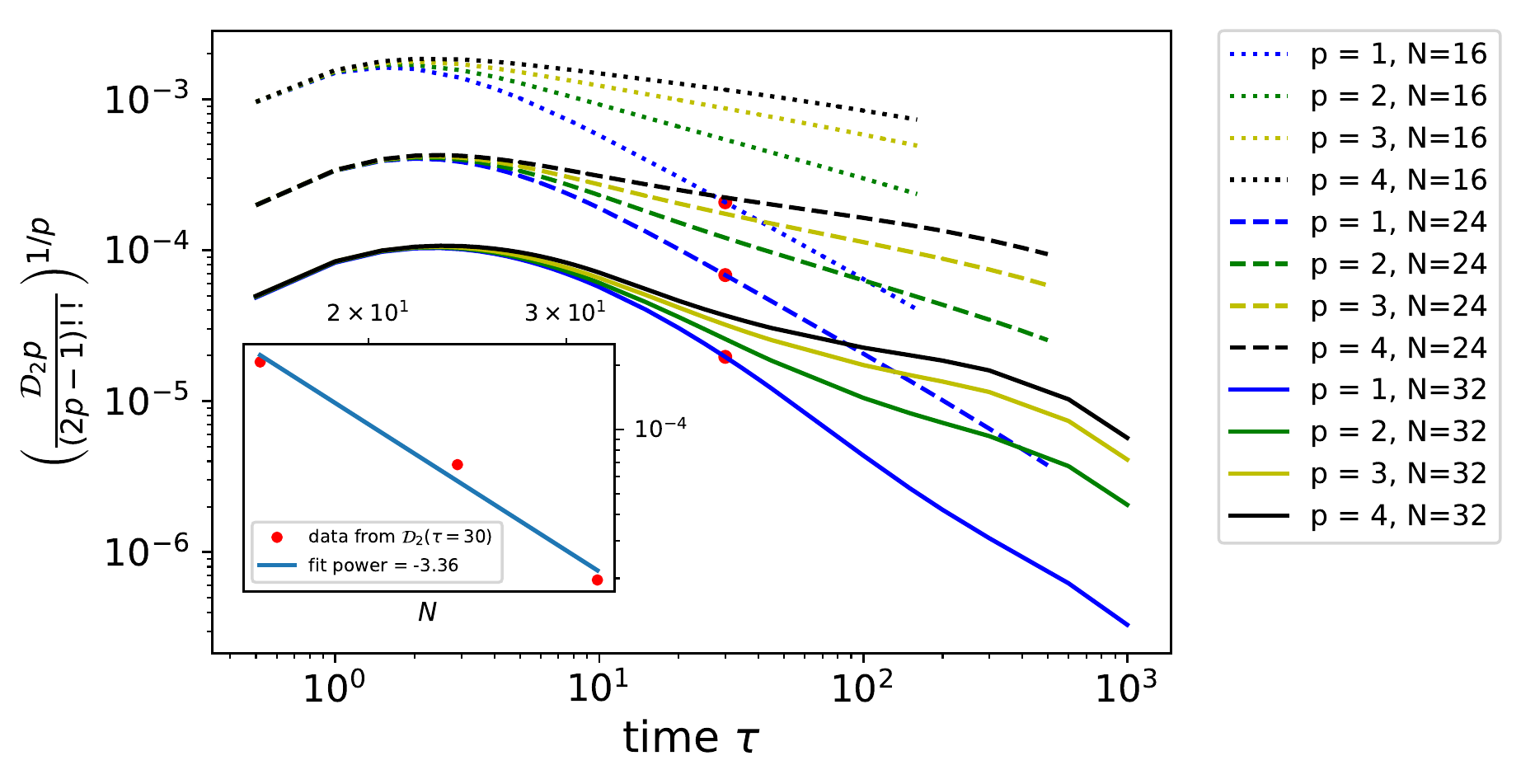}
\caption{Log-Log plot of $\mathcal{D}_{2p}$ versus $\tau$ for $N=16$ (dots), $24$ (dashed) and $32$ (solid), averaged over 50000, 5000 and 30 realizations respectively. Diminishing of the magnitude of correlation functions with $N$ without the change of its time dependence confirms predictions of  fluctuation expansion around the replica diagonal saddle point. Inset: Fit of the amplitude of the correlation function at $\tau=30$ for different $N$. The best fit is achieved for the power $-3.36$, the fluctuation  expansion predicts the power $-3$.}
\label{fig_D2p-N}
\end{figure}

Results of numerical calculations of the dependence $\mathcal{D}_{2p}(\tau)$ are shown in Figs. \ref{fig_D2p-tau} and \ref{fig_D2p-N}. Figure \ref{fig_D2p-tau} shows the time dependence of the correlation function $\mathcal{D}_{2p}(\tau)$ for different  $p$,  for our largest system, $N=32$.  First, one notices the non-monotonous dependence of the correlation functions on time. This short time behavior originates from the fact that equal time expectation $\langle \chi_i(0)\chi_j(0) \rangle=0$ for $i\neq j$ for $N=8,16,24,32,\ldots $, which belong to the orthogonal symmetry class \cite{you2017sachdev}. Indeed, from anticommutation of Majoranas one concludes that $\langle \chi_i\chi_j \rangle$ is pure imaginary. On the other hand, 
for orthogonal symmetry classes, there is a representation of Majorana operators with all matrix elements $\langle n|\chi_i|m\rangle$  being real. This contradiction enforces zero value for equal time expectation. The field theory does not resolve this fact. At longer time, $\tau \gtrsim 3/J$, the correlation functions decrease in time in a qualitative agreement with the field theory. However, while  $p=1$ function is consistent with Eqs. (\ref{D2_RD}), (\ref{D2_RD_rep}), the $p\geq 2$ functions exhibit fast crossover to the two-level regime. We thus are not able to verify 
the time dependence of Eqs.~(\ref{D2p_D2}) even for our largest system of $N=32$ for $p\geq 2$. 

We can, however, verify the $N$-dependence of Eqs.~(\ref{D2p_D2}). Numerical results for the dependence of the correlation functions on the number of sites $N$ are shown in Fig. \ref{fig_D2p-N}. One can see that the  correlations functions rapidly decrease with increasing $N$ while keeping qualitatively the same time dependence. This is in accord with the predictions from the Gaussian fluctuation expansion around the replica-diagonal saddle point. The best fit, see inset in Fig.~\ref{fig_D2p-N}, for power law dependence on $N$ is $N^{-3.36}$, which is close to the power $N^{-3}$,  following from the Gaussian fluctuation expansion. In contrast, the replica off-diagonal saddle point predicts no suppression of the site off-diagonal correlation functions with $N$, Eq.~(\ref{D2p_D2}).  Once again we conclude that the numerics is consistent with the replica-diagonal theory and is inconsistent with the off-diagonal saddle points. 

\section{Conclusions}
\label{sec:conclusions}

We have examined signatures of the replica off-diagonal saddle points in the field theory treatment of the SYK$_4$ model. 
We have argued that such off-diagonal elements affect the coset manifold ${\cal G}/H$ of reparameterization soft modes 
and thus change the expected long-time behavior of ``mesoscopic'' correlation functions. Comparing to numerically   evaluated corresponding correlation functions for $N\leq 32$ SYK$_4$ model, we conclude that they do not show any evidence for replica off-diagonal saddle points. On the contrary, all correlation functions (both site local and non-local) are in a good agreement with the expectations stemming from  the replica  diagonal saddle point plus Gaussian fluctuations. The latter do include replica off-diagonal components, of course. 

We conclude thus that we do not detect any evidence for replica off-diagonal saddle points, at least for time scales shorter than the inverse spacing between the ground-state and the first many-body excited state. At longer times the data is well described by statistics of the two-level system. We stress, though, that such  two-level description is outside of the field theoretical treatment, we base our conclusions at. These conclusions seems to be at odds with the recent proposal for off-diagonal saddles in Refs.~\cite{Stanford_2replicas,Khramtsov}. On the other hand, they are in line with no evidence for glassy behavior in the SYK model reported in   Ref.~\cite{gur2018does}.  

Does it mean that there is no room for replica symmetry breaking and replica off-diagonal structure in SYK-like models?
In our opinion such conclusion is premature. One may investigate deformed models, such as eg. SYK$_4$ + SYK$_2$  \cite{nosaka2018thouless,garcia2018chaotic}.
Our preliminary investigation \cite{tobepublished} points  to similarities between level and eigenfunction statistics of such models to those of random regular graphs (RRG)  \cite{altshuler2016nonergodic,altshuler2016multifractal,KRAVTSOV2018148,Tikhonov2016,Tikhonov2018}. In the case of RRG  replica symmetry breaking was argued to be a proper framework to describe an observed phenomenology \cite{KRAVTSOV2018148}. One is thus justified to expect that phenomenologically similar deformed SYK models may admit similar replica symmetry broken description. However, one should probably conclude that the {\em undeformed} SYK$_4$ model does {\em not} exhibit deviations from the replica diagonal and symmetric saddle point description. 

\acknowledgments
We are grateful to M. Khramtsov, K. Tikhonov and Yi-Ming Wu for useful discussions.  H.W. was supported by DOE contract DEFG02-08ER46482. D.B. was funded by the Deutsche Forschungsgemeinschaft (DFG, German Research Foundation) --- Projektnummer 277101999 ---TRR 183 (project A03). A.K. was supported by NSF grant DMR-1608238. A.C. thanks William I. Fine Theoretical Physics Institute, University of Minnesota for hospitality during the preparation of this work. 

\appendix
\section{Replica non-diagonal saddle point solutions.}
In this section we provide detailed form of the replica non-diagonal solutions of saddle point Eqs. (\ref{MatrixSP}). 
Consider the $n\times n$  matrix ${\bf g}$ with all diagonal elements equal $\tilde{g}$, and all off-diagonal elements equal $g$ 
\begin{equation}
g_{ab}=\tilde{g}\delta_{ab}+g(1-\delta_{ab}), 
\label{g_ab}
\end{equation}
Substituting the ansatz Eq. (\ref{g_ab}) into saddle point Eqs. (\ref{MatrixSP}), we obtain 
\begin{eqnarray}
&& 
\tilde{g}^4+(n-1)g^4=1, \label{g4}\\ 
&& 
\tilde{g}^3g+g^3\tilde{g}+(n-2)g^4=0. \label{g3g}
\end{eqnarray}
Introducing the variable $z=\tilde{g}/g$ we obtain from Eq. (\ref{g3g}) 
\begin{equation}
z^3+z+(n-2)=0. 
\label{Eq-z}
\end{equation}
In the replica-limit $n\rightarrow 0$, equation for $z$ can be written as 
\begin{equation}
(z-1)(z^2+z+2)=0
\label{Eqz_n}
\end{equation} 
The solution $z=1$ means $\tilde{g}=g$, which cannot satisfy Eq. (\ref{g4}). 
The solutions of equation $z^2+z+2=0$,  
\begin{equation}
z=\frac{-1\pm i\sqrt{7}}{2},  
\label{z}
\end{equation}
result in the nontrivial replica non-diagonal saddle-points that are discussed below. 
For $z= \frac{-1 + i\sqrt{7}}{2}$, we obtain 
\begin{equation}
g^4=-\frac{1}{4}e^{i \arctan(3\sqrt{7})}, \, \,  \tilde{g}^4= e^{-i \arctan\left(\frac{3\sqrt{7}}{31}\right)}, 
\label{g4-tilde_g4}
\end{equation}
while $z= \frac{-1 - i\sqrt{7}}{2}$ results in the complex conjugated expressions. It follows, that there are four pairs of mutually complex conjugated solutions with nontrivial replica-off-diagonal part,  $g\neq 0$.  

To determine the relevance of the found saddle point solutions, we compare the values of the saddle-point action at the replica-diagonal and at the replica-off-diagonal saddle points. For the calculation of $\mathrm{Tr}\ln$ term at replica-off-diagonal saddle point we use the following formula 
\begin{eqnarray}
\nonumber && 
\mathrm{Tr}\ln \Sigma^{ab}_{\tau\tau'}= 
\ln\det\left[\hat{\sigma}\right]+ n\mathrm{Tr}\ln\left(\Sigma_{\tau,\tau'}\right)\\ 
\nonumber && 
=\ln\left\{\tilde{\sigma}^n\left(1-\frac{\sigma}{\tilde{\sigma}}\right)^{n-1}\left[1+(n-1)\frac{\sigma}{\tilde{\sigma}}\right]\right\}+ n\mathrm{Tr}\ln\left(\Sigma_{\tau,\tau'}\right). \\
\label{detSigma}
\end{eqnarray}
where $\tilde{\sigma}$ and $\sigma$ denote the diagonal and off-diagonal elements of the replica-matrix $\hat{\sigma}$ for the 
self-energy, and $\Sigma_{\tau,\tau'}$ denotes the replica-diagonal solution as specified in Eqs. (\ref{EqSigmaG-separate}), (\ref{MatrixSP}), (\ref{ansatz_Sigma}). 
Performing the replica limit $n\rightarrow 0$ in Eq. (\ref{detSigma}),  the action for the replica-off-diagonal saddle point  can be written as 
\begin{eqnarray}
\nonumber && 
-S_{\mathrm{RND}}=\frac{Nn}{2}\left\{\ln(J^{3/2}\tilde{g}^3)+\ln\left(1-\frac{1}{z^3}\right)+\right.  \\ 
&& \left. 
\frac{1}{z^3-1}+\frac{3}{4}J^2\tilde{g}^4\left(1-\frac{1}{z^4}\right)+ \mathrm{Tr}\ln\left(\Sigma_{\tau,\tau'}\right)\right\}. 
\label{S_RSB}
\end{eqnarray}
Note that $\frac{3}{4}J^2\tilde{g}^4(1-1/z^4)=3/4$ for both values of $z$ from Eq. (\ref{z}) and all possible solutions in Eq. (\ref{g4-tilde_g4}), as well as for the replica diagonal saddle point $\tilde{g}=1/\sqrt{J}$, $g=0$. Therefore, the difference between the action at the replica-diagonal and at the the replica-non-diagonal saddle points comes from the logarithmic terms only.
It is given by 
\begin{equation}
-(S_{\mathrm{RND}}-S_{\mathrm{RD}})=\frac{Nn}{2}\left\{3\ln(\sqrt{J}\tilde{g})+\ln\left(1-\frac{1}{z^3}\right) +
\frac{1}{z^3-1}\right\}. 
\label{SRND-SRD}
\end{equation}
 The  real part of Eq. (\ref{SRND-SRD})  equals 
$-(S_{\mathrm{RND}}-S_{\mathrm{RD}})=0.0284 \frac{Nn}{2}$ for all solutions listed in Eq. (\ref{g4-tilde_g4}). Since it is positive, the replica-non-diagonal saddle points give the dominant contributions to the replicated partition function $\langle Z^n\rangle=e^{-S}$.  The difference $-(S_{\mathrm{RND}}-S_{\mathrm{RD}})$ for possible solutions from Eq. (\ref{g4-tilde_g4}) is summarized in the table below \\

\begin{tabular}{|c|c|c|c|}
\hline 
z &   $J^2 \tilde{g}$ & $J^2 g$ & $-(S_{\mathrm{RSB}}-S_{\mathrm{RS}})$ \\ 
\hline 
$\frac{-1+i\sqrt{7}}{2}$ & $0.998037 - 0.0626229 i$ & $-0.29093 - 0.644484 i $ & $0.0284264 + 1.66411 i \pm i\frac{3\pi}{2}$ \\ 
\hline 
$\frac{-1+i\sqrt{7}}{2}$ & $0.0626229 + 0.998037 i$ & $0.644484 - 0.29093 i$ & $0.0284264+0.87871 i \pm i\frac{3\pi}{2}$ \\ 
\hline 
$\frac{-1+i\sqrt{7}}{2}$ & $-0.998037 + 0.0626229 i$ & $0.29093 + 0.644484 i$ & $0.0284264+2.44951 i \pm i\frac{3\pi}{2}$ \\ 
\hline 
$\frac{-1+i\sqrt{7}}{2}$ & $-0.0626229 - 0.998037 i$ & $-0.644484 + 0.29093 i$ & $0.0284264-0.692086 i \pm i\frac{3\pi}{2}$ \\ 
\hline 
\end{tabular} \\

\noindent
The values for $z=\frac{-1-i\sqrt{7}}{2}$ are complex conjugated to the corresponding values given in the table.

In conclusion of this section, we consider another possible structure of the non-diagonal saddle point matrix 
 ${\bf g}$, consisting of $n$ blocks of the size $p\times p$ along the main diagonal, the structure of each $p\times p$ block being given by Eq. (\ref{g_ab}).  In that case, Eqs. (\ref{MatrixSP}) result in the following equations for the matrix elements 
\begin{eqnarray}
&& 
\tilde{g}^4+(p-1)g^4=1, \label{g4_p}\\ 
&& 
\tilde{g}^3g+g^3\tilde{g}+(p-2)g^4=0. \label{g3g_p}
\end{eqnarray}
From Eq. (\ref{g3g}) we obtain for $z=\tilde{g}/g$
\begin{equation}
z^3+z+(p-2)=0. 
\label{Eq-z}
\end{equation}
In general, there are three solutions for $z$. However, only two mutually complex conjugated solutions are consistent with Eq. (\ref{g4}). Each solution fixes unambiguously the values of $\tilde{g}^4$ and $g^4$, thus resulting in four possible (generally complex) values of $\tilde{g}$ and $g$. In particular, for $p=2$ we obtain 
\begin{equation}
z=0, \, \, \, {\mbox or} \, \, \, z=\pm i. 
\label{z_p2}
\end{equation}
The solution $z=0$ results in the completely replica off-diagonal matrix ${\bf g}$, which clearly contradicts numerical results in Sec. \ref{secSite-local}. From the solution $z=i$ it follows $\tilde{g}=i g$. Substituting this relation in Eq. (\ref{g4}) for $p=2$, we obtain, as one possible solution, the matrix ${\bf g}$ as given in Eq. (\ref{22MatrixSP}).

\section{Fluctuation expansion around the replica-diagonal saddle point}
To obtain the action for massive fluctuations around the replica diagonal saddle point, we adopt the ansatz 
\begin{eqnarray}
&& 
G^{ab}_{\tau\tau'}=G(\tau-\tau')\delta_{ab} +\delta G^{ab}_{\tau\tau'}, 
\label{deltaG} \\
&& 
\Sigma^{ab}_{\tau\tau'}=\Sigma(\tau-\tau')\delta_{ab}+\delta\Sigma^{ab}_{\tau\tau'}, 
\label{deltaSigma}  
\end{eqnarray} 
where $G(\tau-\tau')$ and $\Sigma(\tau-\tau')$ denote the traditional replica-diagonal saddle point solutions. Substituting Eqs. (\ref{deltaG}), (\ref{deltaSigma})  into the action Eq. (\ref{SYC-Action}) and performing the expansion in $\delta\Sigma$ and $\delta G$, we obtain 
the action for the  fluctuations around the the replica-diagonal saddle point, which can be represented as a sum of the actions for different replicas, 
\begin{equation}
-\delta S[\delta\Sigma, \delta G]=-\sum_{a,b=1}^n \delta S^{ab}[\delta\Sigma^{ab}, \delta G^{ab}], 
\label{Sum_Sab}
\end{equation}
where 
\begin{eqnarray}
\nonumber && 
\delta S^{aa}[\delta\Sigma^{aa}, \delta G^{aa}]=\frac{N}{2}\left\{\frac{1}{2}\mathrm{Tr}\left[G\delta\Sigma^{aa}G\delta\Sigma^{aa}\right]- \int d\tau d\tau' \delta\Sigma^{aa}_{\tau'\tau}\delta G^{aa}_{\tau\tau'} -\right. \\ 
\nonumber && 
\left. 
\frac{1}{4}J^2 \int d\tau d\tau' \left[6 G^2(\tau-\tau')\left(\delta G^{aa}_{\tau\tau'}\right)^2+
4 G(\tau-\tau')\left(\delta G^{aa}_{\tau\tau'}\right)^3+\left(\delta G^{aa}_{\tau\tau'}\right)^4\right]
\right\} \\ 
\label{delta_Saa}
\end{eqnarray}
\begin{equation} 
\delta S^{ab}[\delta\Sigma^{ab}, \delta G^{ab}]=
\frac{N}{2}\left\{\frac{1}{2}\mathrm{Tr}\left[G\delta\Sigma^{ab}G\delta\Sigma^{ba}\right]- 
\int d\tau d\tau' \delta\Sigma^{ba}_{\tau'\tau}\delta G^{ab}_{\tau\tau'} -\frac{1}{4}J^2\left(\delta G^{ab}_{\tau\tau'}\right)^4 \right\}. 
\label{delta_Sab}
\end{equation}
Let us derive the generating functional for the correlation functions containing off-diagonal fluctuations $\delta\Sigma^{ab}$.  To this end we extend the action Eq. (\ref{delta_Sab}) with source terms 
 for $\delta\Sigma$ and $\delta G$
\begin{equation}
-S_{\mathrm{source}}=\frac{N}{2}\int d\tau d\tau' \left\{j^{ba}_{\tau'\tau}\delta G^{ab}_{\tau\tau'}+\delta\Sigma^{ba}_{\tau'\tau}h^{ab}_{\tau\tau'}\right\}
\label{Ssource}
\end{equation}
The quadratic part of the action $-S=-\delta S^{ab}[\delta\Sigma^{ab}, \delta G^{ab}]-S_{\mathrm{source}}$  can be represented in the matrix form 
\begin{eqnarray}
\nonumber && 
-\delta S^{ab}_{(2)}= \frac{1}{2}(h^{ab}_{\tau\tau'},j^{ab}_{\tau\tau'})\cdot \left(
\begin{array}{c}
\delta\Sigma^{ba}_{\tau',\tau} \\  
\delta G^{ba}_{\tau'\tau}
\end{array}
\right)  -\\ 
 &&   
\frac{N}{4}\left(\delta\Sigma^{ab}_{\tau_1,\tau_2}, \delta G^{ab}_{\tau_1,\tau_2}\right) 
\left(\begin{array}{cc}
\mathcal{K}^{\tau_2,\tau_3}_{\tau_1,\tau_4} & -\delta_{\tau_1,\tau_4}\delta_{\tau_2,\tau_3} \\
-\delta_{\tau_1,\tau_4}\delta_{\tau_2,\tau_3} & 0 
\end{array}
\right)
\left(\begin{array}{c}
\delta\Sigma^{ba}_{\tau_3,\tau_4} \\  
\delta G^{ba}_{\tau_3\tau_4}
\end{array}
\right) 
\end{eqnarray}
Using the symmetry of the fermionic Green functions with respect to exchange of arguments 
\begin{equation}
\delta G^{ab}_{\tau_1,\tau_2}=-\delta G^{ba}_{\tau_2,\tau_1}, \, \, \, 
\delta\Sigma^{ab}_{\tau_1,\tau_2}=-\delta\Sigma^{ba}_{\tau_2,\tau_1}
\end{equation}
we rewrite the quadratic action in the form 
\begin{eqnarray}
-\delta S^{ab}_{(2)} &=& \frac{N}{4}\left(\delta\Sigma^{ab}_{\tau_1,\tau_2}, \delta G^{ab}_{\tau_1,\tau_2}\right) 
\left(\begin{array}{cc}
\mathcal{K}^{\tau_2,\tau_4}_{\tau_1,\tau_3} & -\delta_{\tau_1,\tau_3}\delta_{\tau_2,\tau_4} \\
-\delta_{\tau_1,\tau_3}\delta_{\tau_2,\tau_4} & 0 
\end{array}
\right)
\left(\begin{array}{c}
\delta\Sigma^{ab}_{\tau_3\tau_4} \\  
\delta G^{ab}_{\tau_3\tau_4}
\end{array} 
\right)  \nonumber \\
&-& \frac{1}{2}(h^{ab}_{\tau\tau'},j^{ab}_{\tau\tau'})\cdot \left(
\begin{array}{c}
\delta\Sigma^{ab}_{\tau,\tau'} \\  
\delta G^{ab}_{\tau\tau'}
\end{array}
\right)  
\label{quadAction}
\end{eqnarray}
Correlation functions of the fields $\delta G$ and $\delta\Sigma$ are obtained by taking derivatives, for instance 
\begin{equation}
\langle\delta\Sigma^{ab}_{\tau_1\tau_2}\delta\Sigma^{ab}_{\tau_3\tau_4}(\delta G^{ab}_{\tau_5,\tau_6})^4\rangle= 
2^6 \frac{\delta^6\langle e^{-\delta S^{ab}_{(2)}}\rangle}{\delta h^{ab}_{\tau_1\tau_2} 
\delta h^{ab}_{\tau_3\tau_4} (\delta j^{ab}_{\tau_5\tau_6})^4 }
\end{equation}
Integrating the fields $\delta\Sigma$, $\delta G$, we obtain 
\begin{equation}
\langle e^{-S^{ab}[j, h]}\rangle = \exp\left[\frac{1}{4N}(h^{ab}_{\tau_1\tau_2},j^{ab}_{\tau_1\tau_2})\left(\begin{array}{cc}
0 & \delta_{\tau_1,\tau_3}\delta_{\tau_2,\tau_4} \\
\delta_{\tau_1,\tau_3}\delta_{\tau_2,\tau_4} & \mathcal{K}^{\tau_2,\tau_4}_{\tau_1,\tau_3} 
\end{array}
\right)
\left(\begin{array}{c}
h^{ab}_{\tau_3\tau_4} \\  
j^{ab}_{\tau_3\tau_4}
\end{array}
\right)\right]
\label{genFun}
\end{equation}
From Eq. (\ref{genFun}) we read off the following nonzero contractions of the fields $\delta G$ and $\delta\Sigma$ 
\begin{eqnarray}
&& 
\langle\delta G^{ab}_{\tau_1,\tau_2} \delta \Sigma^{ab}_{\tau_3,\tau_4} \rangle= \langle\delta \Sigma^{ab}_{\tau_1,\tau_2} \delta G^{ab}_{\tau_3,\tau_4} \rangle= \frac{1}{4N}\delta_{\tau_1,\tau_3}\delta_{\tau_2,\tau_4}, \\ 
&& 
\langle\delta G^{ab}_{\tau_1,\tau_2} \delta G^{ab}_{\tau_3,\tau_4} \rangle=\frac{1}{4N}\mathcal{K}^{\tau_2,\tau_4}_{\tau_1,\tau_3}= 
\frac{1}{4N}G(\tau_3-\tau_1) G(\tau_2-\tau_4). 
\label{Contractions}
\end{eqnarray}
The additive form of the fluctuation action Eq. (\ref{Sum_Sab}) implies the multiplicative form of the correlation function Eq. (\ref{D2p_replica}) as a product over the pairs of replicas 
\begin{eqnarray}
\nonumber && 
\mathcal{D}_{2p}(\tau)=(2p-1)!! \langle\chi^{a_1}_i(\tau)\chi^{a_2}_i(0)\rangle\langle\chi^{a_1}_j(0)\chi^{a_2}_j(\tau)\rangle...\langle \chi^{a_{2p-1}}_i(\tau)\chi^{a_{2p}}_i(0)\rangle\langle\chi^{a_{2p-1}}_j(0)\chi^{a_{2p}}_j(\tau)\rangle \\ 
&& 
= (2p-1)!! \left(\mathcal{D}_{2}(\tau)\right)^p. 
\label{D2p_product}
\end{eqnarray}
Here $(2p-1)!!$ denotes the number of ways to assign pairs of replicas. 
 The fluctuation expansion around the replica-diagonal saddle point results in 
\begin{equation}
D^{\mathrm{RD}}_2(\tau)= \frac{3}{2^{10}N^3}G(\tau)G(-\tau)= \frac{3}{4\pi 2^{10}N^3}\frac{1}{|\tau|}.
\label{D2_RD-appendix}
\end{equation} 
\begin{figure}[ht]
\includegraphics[width=\linewidth]{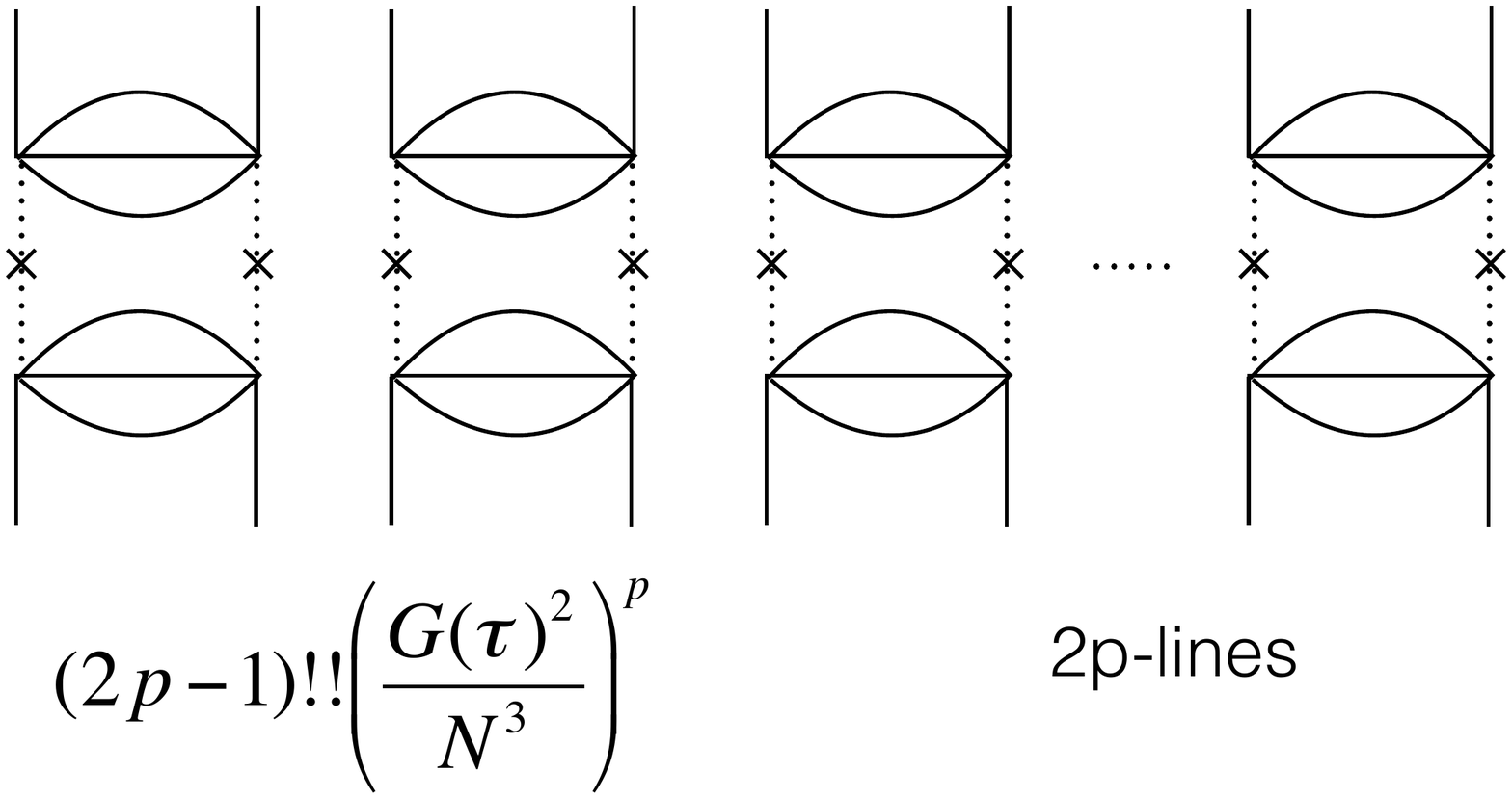}
\vskip -3cm
\caption{Diagram for the correlation function  $\mathcal{D}_{2p}$. Crossed dashed lines represent averaging over disorder.}
\centering
\label{diagram-D2p}
\end{figure}
The same conclusion can be made by means of $1/N$ diagrammatic expansion without employing the replica-trick (see Fig. \ref{diagram-D2p}) \cite{Polchinski-Rosenhaus2016}.

\section{Correlation function in the replica non-diagonal saddle point.}

The replica non-diagonal saddle point assigns a non-zero average to the correlation function $\langle\chi_i^{a}(\tau)\chi_i^{b}(\tau')\rangle$. Calculation of high-order correlation functions, such as the one defined by Eq. (\ref{D2p}), requires building of contractions of pairs of fermions corresponding to the nonzero average in the saddle point, which leads to the general form as given by Eq. (\ref {D2p_product}), but with the function $\mathcal{D}_2$ calculated in the replica non-diagonal saddle point.  

Let us now consider the calculation of the function $\mathcal{D}_2$ explicitely. First we note, that the saddle point equations Eqs. (\ref{EqSigmaG}) are invariant with respect to the replica dependent time shift $\tau\rightarrow \tau+\tau_a$. This transformation does not change the time dependence in the replica-diagonal elements of saddle point solutions in Eq.  (\ref{ansatz_Sigma}), it influences the replica off-diagonal elements though. As we show below, the time-shift transformation is crucial to obtain the correct time dependence for the correlation functions calculated at the replica non-diagonal saddle point. 
Consider the correlation function between Majorana fermions on different sites ($i\neq j$)
\begin{equation}
D_{2}(\tau_1, \tau_2; \tau_3, \tau_4) = \left\langle \langle\chi_i(\tau_1)\chi_j(\tau_2)\rangle_{\mathrm{QM}}
 \langle\chi_i(\tau_3)\chi_j(\tau_4)\rangle_{\mathrm{QM}}\right\rangle_{\mathrm{dis}}. 
\label{D2ij}
\end{equation}
The correlation function Eq. (\ref{D2ij}) in the replica formalism reads 
\begin{equation}
D_{2}(\tau_1, \tau_2; \tau_3, \tau_4)=\left\langle \chi^a_i(\tau_1)\chi^a_j(\tau_2)
\chi^b_i(\tau_3)\chi^b_j(\tau_4)\right\rangle=-\left\langle G^{ab}(\tau_1-\tau_3) G^{ab}(\tau_2-\tau_4)\right\rangle_{\Phi},
\label{D2-replicas}
\end{equation}
where $\langle ... \rangle_{\Phi}$ denotes the average over the reparametrization fluctuations, and 
$G^{ab}(\tau-\tau')$ denotes the saddle point replica-off-diagonal Green's function. The time-dependence in Eq. (\ref{D2-replicas}) contradicts the quantum mechanical result, which predicts the dependence of correlation function on the differences $\tau_1-\tau_2$ and $\tau_3-\tau_4$. However, one can restore the correct quantum mechanical dependence of the correlation function Eq. (\ref{D2-replicas}) by making appropriate time-shifts in each replica. Namely, for each pair of times belonging to the same replica, the time-shift has to be chosen in such a way, that the shifted times are symmetric with respect to zero.  Specifically, in Eq. (\ref{D2-replicas}), the times $\tau_1, \tau_2$, belonging to the replica $a$, are shifted by $c_a=-(\tau_1+\tau_2)/2$, so that after the shift $\tau'_1=\tau_1-c_a=(\tau_1-\tau_2)/2$, and $\tau'_2=\tau_2-c_a=-(\tau_1-\tau_2)/2=-\tau'_1$. Correspondingly, the times $\tau_3$ and $\tau_4$ are shifted by $c_b=-(\tau_3+\tau_4)/2$. Let us calculate the correlation function Eq. (\ref{D2-replicas}) assuming the following ordering of times: $\tau_2<\tau_3<\tau_4<\tau_1$. Using the Liouville quantum mechanical representation for the averaging over the reparametrization fluctuations \cite{Bagrets-Altland-Kamenev2016}, we obtain 
\begin{eqnarray}
\nonumber && 
\left\langle G^{ab}(\tau_1-\tau_3)G^{ab}(\tau_2-\tau_)\right\rangle_{\Phi} \propto 
\int_0^{\infty} \frac{d\alpha d\beta}{\sqrt{\alpha\beta}} \sum_{k}\langle 0|e^{\frac{1}{4}\phi}|k_{\alpha}\rangle 
e^{-\frac{k_{\alpha}^2}{4M}[(\tau_3-\tau_2)+c_b-c_a)]}\times \\ 
\nonumber && 
\langle k_{\alpha}|e^{\frac{1}{4}\phi}|k_{\alpha+\beta}\rangle e^{-\frac{k_{\alpha+\beta}^2}{2M}(\tau_4-\tau_3)}
\langle k_{\alpha+\beta}|e^{\frac{1}{4}\phi}|k_{\beta}\rangle e^{-\frac{k_{\beta}^2}{4M}[(\tau_1-\tau_3)+c_a-c_b}
\langle k_{\beta}|e^{\frac{1}{4}\phi}|0\rangle =\\ 
\nonumber &&  
\int_0^{\infty} \frac{d\alpha d\beta}{\sqrt{\alpha\beta}} \sum_{k}\langle 0|e^{\frac{1}{4}\phi}|k_{\alpha}\rangle 
e^{-\frac{k_{\alpha}^2}{4M}[\frac{1}{2}(\tau_3-\tau_4+\tau_1-\tau_2)]}\times \\ 
\nonumber && 
\langle k_{\alpha}|e^{\frac{1}{4}\phi}|k_{\alpha+\beta}\rangle e^{-\frac{k_{\alpha+\beta}^2}{2M}(\tau_4-\tau_3)}
\langle k_{\alpha+\beta}|e^{\frac{1}{4}\phi}|k_{\beta}\rangle e^{-\frac{k_{\beta}^2}{4M}[\frac{1}{2}(\tau_1-\tau_2+\tau_4-\tau_3)}
\langle k_{\beta}|e^{\frac{1}{4}\phi}|0\rangle 
\label{GabGab}
\end{eqnarray}
One can see, that after the time shifts $c_a$ and $c_b$, the correlation function depends on the time differences $\tau_1-\tau_2$ and $\tau_4-\tau_3$ only.  
For $\tau_1=\tau_4=\tau$, $\tau_2=\tau_3=0$ Eq. (\ref{GabGab}) reduces to the correlation function calculated in Ref. \cite{Bagrets-Altland-Kamenev2016}, namely 
\begin{eqnarray}
\nonumber && 
\left\langle G^{ab}(\tau)G^{ab}(\tau)\right\rangle_{\Phi} \propto 
\int_0^{\infty} \frac{d\alpha d\beta}{\sqrt{\alpha\beta}} \sum_{k}\langle 0|e^{\frac{1}{2}\phi}|k_{\alpha+\beta}\rangle e^{-\frac{k_{\alpha+\beta}^2}{2M}\tau}
\langle k_{\alpha+\beta}|e^{\frac{1}{2}\phi}|0\rangle \sim \frac{1}{\tau^{3/2}}. 
\label{GabGab_tau}
\end{eqnarray}

Let us now consider the higher powers of the site-nonlocal correlation functions, which we define as a product of $2p$ quantum mechanical averages 
\begin{eqnarray}
\nonumber && 
\mathcal{K}(\tau_1, ... , \tau_{4p})=\left\langle\langle \chi_i(\tau_1)\chi_j(\tau_2)\rangle_{\mathrm{QM}} \langle \chi_i(\tau_3)\chi_j(\tau_4)\rangle_{\mathrm{QM}}... \langle \chi_i(\tau_{4p-1})\chi_j(\tau_{4p})\rangle_{\mathrm{QM}}\right\rangle_{\mathrm{dis}}=\\ 
&& 
\left\langle \chi_i^{a_1}(\tau_1)\chi_i^{a_2}(\tau_3)...  \chi_i^{a_{2p}}(\tau_{4p-1}) 
\chi_j^{a_1}(\tau_2)\chi_j^{a_2}(\tau_4)...  \chi_j^{a_{2p}}(\tau_{4p}) \right\rangle
\label{power_2p}
\end{eqnarray}
To facilitate the transition to the limit case, considered in Section \ref{sec_Site-non-local} 
\begin{eqnarray}
\nonumber && 
\mathcal{D}_{2p}(\tau)=\left\langle\left(\langle \chi_i(\tau)\chi_j(0)\rangle_{\mathrm{QM}}\langle \chi_j(\tau)\chi_i(0)\rangle_{\mathrm{QM}}\right)^p\right\rangle_{\mathrm{dis}}=\\ 
&& 
(-1)^p \left\langle\left(\langle \chi_i(\tau)\chi_j(0)\rangle_{\mathrm{QM}}\langle \chi_i(0)\chi_j(\tau)\rangle_{\mathrm{QM}}\right)^p\right\rangle_{\mathrm{dis}}
\label{def_D2p}
\end{eqnarray}
we adopt the following ordering of times in Eq. (\ref{power_2p})
\begin{eqnarray}
\nonumber && 
\tau_2<\tau_3<\tau_6<\tau_7<...<\tau_{4p-2}<\tau_{4p-1}<\tau_{4p}<\tau_{4p-3}<\tau_{4p-4}<\tau_{4p-7}<\tau_{4p-8}<... \\ 
&& 
<\tau_5<\tau_4<\tau_1
\label{time-ordering_ij}
\end{eqnarray}
The transition to Eq. (\ref{def_D2p}) is achieved by taking the limits $\tau_2=\tau_3=...=\tau_{4p}=0$, 
$\tau_{4p-3}=\tau_{4p-4}=...=\tau_1=\tau$. 
The correlation function Eq. (\ref{power_2p}) is given by the sum over all possible site-local contractions between the fermions belonging to different replicas. Consider a single contribution, where we denote the pairs of contracted replicas as  $(a_1, a_2)$, $(a_3, a_4)$, ... $(a_{2p-1}, a_{2p})$. Then, employing the Liouville quantum mechanincal treatment of the averaging over reparametrizations and implementing replica-dependent time shifts $\tau_a\rightarrow \tau_a+c_a$, we obtain 
\begin{eqnarray}
\nonumber && 
\mathcal{K}(\tau_1, ... , \tau_{4p})=\left\langle G^{a_1, a_2}(\tau_1, \tau_3) G^{a_1, a_2}(\tau_2, \tau_4) 
G^{a_3, a_4}(\tau_5, \tau_7) G^{a_3, a_4}(\tau_6, \tau_8)... \right. \\ 
\nonumber && 
\left. 
 G^{a_{2p-1}, a_{2p}}(\tau_{4p-3}, \tau_{4p-1})
G^{a_{2p-1}, a_{2p}}(\tau_{4p-2}, \tau_{4p})\right\rangle_{\Phi}\propto \\ 
\nonumber && 
\int_0^{\infty} \frac{d\alpha_1 ... d\alpha_{2p}}{\left(\alpha_1 ... \alpha_{2p}\right)^{1/2}} 
\sum_{k_{\alpha}, k'_{\alpha}} \langle 0|e^{\frac{\phi}{4}}|k_{\alpha_1}\rangle 
\exp\left[-\frac{k^2_{\alpha_1}}{2M}(\tau_3+c_2-\tau_2-c_1)\right] \\ 
\nonumber && 
\langle  k_{\alpha_1}|e^{\frac{\phi}{4}}|k_{\alpha_1+\alpha_2} \rangle
\exp\left[-\frac{k^2_{\alpha_1+\alpha_2}}{2M}(\tau_7+c_4-\tau_6-c_3)\right]... \\ 
\nonumber && 
\langle  k_{\alpha_1+\alpha_2+...+\alpha_{2p-1}}|e^{\frac{\phi}{4}}|k_{\alpha_1+\alpha_2+...+\alpha_{2p}}\rangle 
\exp\left[-\frac{k^2_{\alpha_1+\alpha_2+...+\alpha_{2p}}}{2M}(\tau_{4p}-\tau_{4p-1})\right] \\ 
\nonumber && 
\langle  k_{\alpha_1+\alpha_2+...+\alpha_{2p}}|e^{\frac{\phi}{4}}|k'_{\alpha_1+\alpha_2+...+\alpha_{2p-1}}\rangle 
\exp\left[-\frac{k'^2_{\alpha_1+\alpha_2+...+\alpha_{2p-1}}}{2M}(\tau_{4p-3}+c_{2p-1}-\tau_{4p}-c_{2p})\right] \\
\nonumber && 
 \langle k'_{\alpha_1+\alpha_2+...+\alpha_{2p-1}}| e^{\frac{\phi}{4}}|k'_{\alpha_1+\alpha_2+...+\alpha_{2p-2}} \rangle 
\exp\left[-\frac{k'^2_{\alpha_1+\alpha_2+...+\alpha_{2p-2}}}{2M}(\tau_{4p-7}+c_{2p-3}-\tau_{4p-4}-c_{2p-2})\right]... \\ 
\nonumber && 
\langle k'_{\alpha_1+\alpha_2+\alpha_{3}} | e^{\frac{\phi}{4}}|k'_{\alpha_1+\alpha_2}\rangle 
\exp\left[-\frac{k'^2_{\alpha_1+\alpha_2}}{2M}(\tau_{4}+c_{2}-\tau_{5}-c_{3})\right] \\ 
&& 
\langle k'_{\alpha_1+\alpha_2} | e^{\frac{\phi}{4}}|k'_{\alpha_1}\rangle 
\exp\left[-\frac{k'^2_{\alpha_1}}{2M}(\tau_{1}+c_{1}-\tau_{4}-c_{2})\right]
\langle k'_{\alpha_1} | e^{\frac{\phi}{4}}|0 \rangle. 
\label{K_QM}
\end{eqnarray} 
To ensure the dependence of Eq. (\ref{K_QM}) on the differences of times belonging to the same replica only, we choose the time shifts $c_k$ as follows 
\begin{equation}
c_k=-\frac{1}{2}(\tau_{2k}+\tau_{2k-1})
\label{choice_c-powerp}
\end{equation}
With this choice, the combinations of times entering the exponents in Eq. (\ref{K_QM}) become 
\begin{eqnarray}
\nonumber && 
\tau_{2k-2}+c_{k-1}-\tau_{2k}-c_k=\frac{1}{2}[(\tau_{2k-2}-\tau_{2k-3})-(\tau_{2k}-\tau_{2k-1}), \\
&& 
\tau_{2k-1}+c_{k}-\tau_{2k-3}-c_{k-1}= \frac{1}{2}[(\tau_{2k-1}-\tau_{2k})-(\tau_{2k-3}-\tau_{2k-2}).
\label{shifted_times}
\end{eqnarray} 
Therefore, the choice of time shifts in Eq. (\ref{shifted_times}) makes the argument of each exponent in Eq. (\ref{K_QM}) to depend only on differences of times in the same replica. Note furthermore, that the time-shifts introduced by Eq. (\ref{choice_c-powerp}) are replica-local, hence they ensure the quantum mechanically correct time dependence for any choice of contractions. 

In the limit $\tau_2=\tau_3=...=\tau_{4p}=0$, $\tau_{4p-3}=\tau_{4p-4}=...=\tau_1=\tau$, the calculation in Eq. (\ref{K_QM}) reduces literally to the one performed in Ref. \cite{Bagrets-Altland-Kamenev2016}, hence one obtains for the correlation function Eq. (\ref{def_D2p}) the time dependence $\sim 1/|\tau|^{3/2}$ for any power $p$.

\section{Details of numerical simulations}
In this paper, we use exact diagonalization to investigate the Green's function of SYK model. Majorana fermion operators are represented by $\gamma$ - matrices,  which can be constructed by Clifford algebra \cite{Garcia16}. Let $\gamma_i$ to be the representation of the operator $\chi_i$. Then one can define the $\gamma_i^{(N)}$ iteratively. When $N=2$, one have $\gamma$ - matrices as following
\begin{equation}
	\gamma^{(2)}_1 = \sigma_1, \quad \gamma^{(2)}_2 = \sigma_2, \quad \gamma^{(2)}_3 = \sigma_3,
\end{equation}
where $\sigma_1,\sigma_2,\sigma_3$ are the Pauli matrices. Assume we have got $\gamma_i^{(d)}$, then we define
\begin{equation}
	\begin{aligned}
	& \gamma_i^{(d+2)} = \sigma_1 \otimes \gamma_i^{(d)}, \, i=1,...,d+1
	\\ & \gamma_{d+2}^{(d+2)} = \sigma_2 \otimes \mathbbm{1}_{2^{d/2}}
	\\ & \gamma_{d+3}^{(d+2)} = \sigma_3 \otimes \mathbbm{1}_{2^{d/2}}
	\end{aligned}
\end{equation}
Diagonalizing the SYK Hamiltonian, one can get all the eigenvalues and eigenstates to construct quantities we want to learn.

\section{Reparametrization fluctuations around the replica non-diagonal saddle point}
In this Appendix we show, that the replica non-diagonal saddle point generates coupling between the reparametrizations in different replicas, leaving only a single soft mode, in which all replicas have the same reparametrization. 
Consider the replica non-diagonal saddle point given by Eq.  (\ref{ansatz_Sigma}), and consider the replica off-diagonal part of the soft-mode action
\begin{equation}
\label{eq:S_f}
S_2[f] =  {N\over 4}\, \mathrm{Tr} (\partial_{\tau} G\partial_{\tau} G) = {N\over 4}\, 
\sum_{ab}\int\limits_{|\tau_1 - \tau_2|>1/J} d\tau_1d\tau_2 \, \partial_{\tau_1}
\left( G^{ab}[f]_{\tau_1,\tau_2} \right) \partial_{\tau_2}\left( G^{bb}[f]_{\tau_2,\tau_1} \right),
\end{equation}
which is formulated in terms of the  reparametrized Green functions
\begin{equation}
\label{eq:Gr}
G^{ab}[f]_{\tau_1,\tau_2} = f'_a(\tau_1)^{1/4} G^{ab}
\Bigl[f_a(\tau_1) -f_b(\tau_2) \Bigr]f_b'(\tau_2)^{1/4},
\end{equation} 
with $f_a(\tau)$ being the reparametrization transformation in the replica $a$.
Note, that such form of the Green's function is valid only for times $|\tau-\tau'| > 1/J$.
For shorter times, $|\tau-\tau'| \ll  1/J$, the Green's function in the model
with finite strength of interaction $J$ should approach a free Majorana correlator, 
$G_{\rm free}^{ab}(\tau) = -\delta^{ab}{\rm sgn}(\tau)/|\tau|$.
Therefore we restricted the domain of integration in Eq.~(\ref{eq:S_f}).

If one further changes the time-integration variables to $t_i = f(\tau_i)$, defines the field
$\zeta^a_t = [(f_a^{-1})'(t)]^{-1/2}$ and integrates by parts then the action $S_2[f]$ can be cast in the following form (for details see Ref. \cite{Bagrets-Altland-Kamenev2016})
\begin{equation}
S_2[f] =\sum_{ab}  \int \!d t_1\, d t_2\, \zeta^a_{t_1}\, \Pi^{ab}(t)\, \zeta^b_{t_2}, 
\qquad \Pi^{ab}(t_1-t_2) =  - 
{N\over 4}\, G^{ab}(t_1-t_2) \overleftrightarrow{\partial_{t_1}} \overleftrightarrow{\partial_{t_2}} 
G^{ab}(t_2-t_1),
\end{equation}
where we introduced  $f_1(t) \overleftrightarrow{\partial_{t}} f_2(t) \equiv \frac 12 [f_1(t) f_2'(t) - f_1'(t) f_2(t)] $ for any two functions $f_1$ and $f_2$. Taking into account the symmetries of the Green's function,
\begin{equation}
G^{ab}(-t) = - G^{ab}(t), \qquad G^{ab} (t)=  G^{ba}(t),
\end{equation}
the polarization operator can be represented in the equivalent form,
\begin{equation}
\label{eq:Pi}
\Pi^{ab}(t) = {N\over 8} \Bigl(  
[ \partial_t G^{ab}(t)]^2   - G^{ab}(t) \partial_t^2 G^{ab} (t) \Bigr).
\end{equation}
This expression needs to be found only for times $|\tau-\tau'| > 1/J$
and therefore one can omit the action of time derivative on the sign-function in Eq.  (\ref{ansatz_Sigma}).
Indeed, the resulting $\delta$-function will
bring times $t_{1,2}$ infinitely close to each other, but these times are excluded 
from the integration domain in Eq.~(\ref{eq:S_f}). Bearing this remark one obtains
\begin{equation}
\Pi^{ab}(t) = - \frac{N}{32\sqrt{\pi} J} \frac{g_{ab}^2}{|t|^3},
\end{equation} 
with $g_{aa}=\tilde{g}$ and $g_{ab}=g$ for $a\neq b$  according to Eq. (\ref{g_ab}). 
Then the soft-mode action assumes the form
\begin{eqnarray}
     \nonumber       
S_2[f]=- \frac{N}{32\sqrt{\pi} J} \sum_{ab} g_{ab}^2\int\!\!\!\!\int d t_1 d t_2 \,
\frac{\zeta^a_{t_1} \zeta^b_{t_2}}{|t_1-t_2|^3}  &&  \\ 
\overset{t_i = f(\tau_i)}{=} - \frac{N}{32\sqrt{\pi} J} \sum_{ab} g_{ab}^2\int\!\!\!\!\int d\tau_1 d\tau_2 \,
\frac{f_a'(\tau_1)^{3/2} f_b'(\tau_2)^{3/2}}{ |f_a(\tau_1)-f_b(\tau_2)|^3}. && 
\label{eq:S2_reg}
\end{eqnarray}
Diagonal matrix elements in this sum produce the Schwarzian action for each function $f_a(\tau)$,  
see e.g. derivation in Ref.~\cite{Bagrets-Altland-Kamenev2016}, and below we analyze the terms with $a \neq b$.
\begin{figure}[t]
\begin{center}
\includegraphics[width=5.0cm]{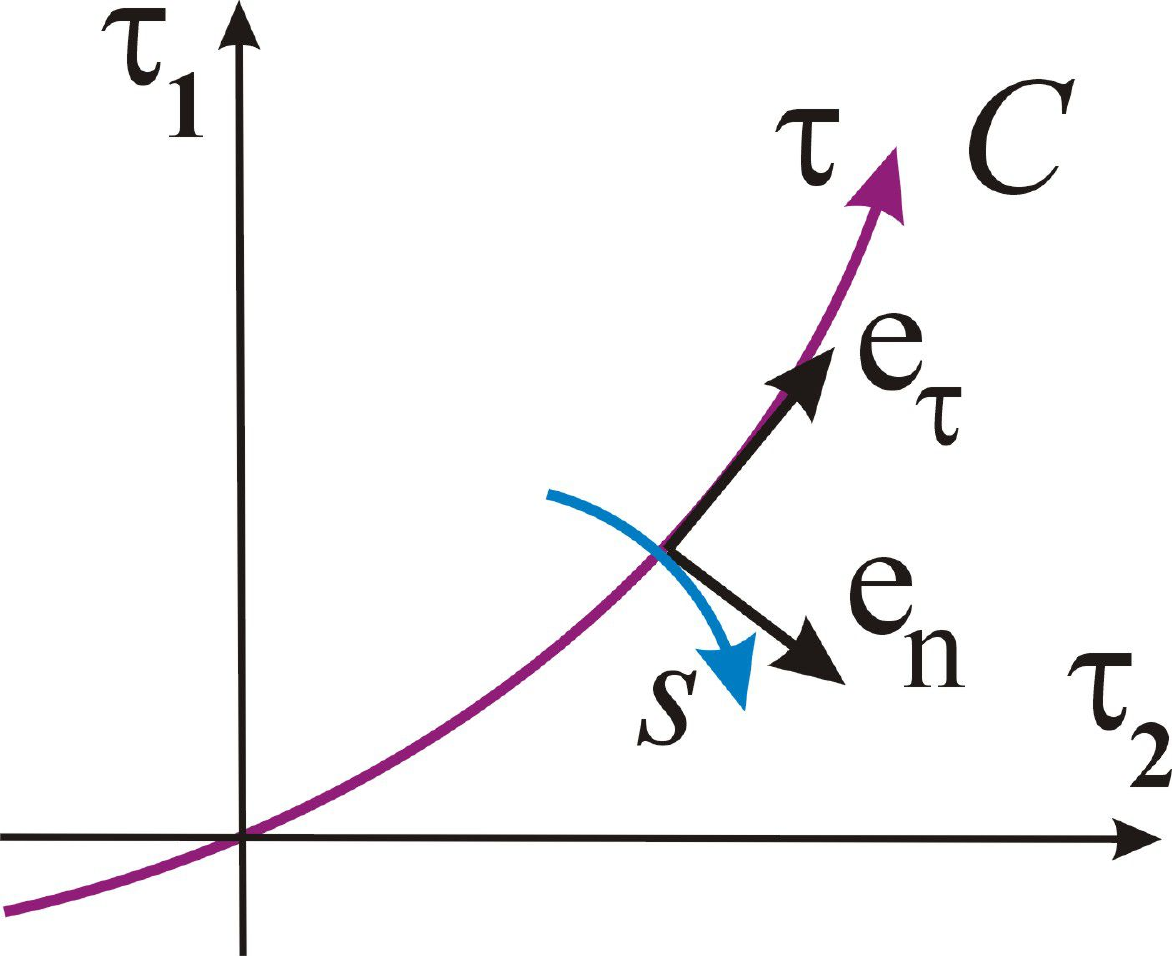}
\caption{The curve ${\cal C}$ is defined by the relation $f_a(\tau_1) = f_b(\tau_2)$. The new (local) system of coordinates
$(\tau,s)$ is chosen as described in the main text so that $\tau$ runs along ${\cal C}$ and $s$ in the orthogonal direction.
The main contribution to the action~(\ref{eq:S2_reg}) stems from the region $|s| \lesssim 1/J$.}
\label{fig:C}
\vskip -1cm
\end{center}
\end{figure}
These off-diagonal contributions mainly come from the singularity along the curve ${\cal C}$ 
in the plane $(\tau_1, \tau_2)$ defined by the equation $f_a(\tau_1)=f_b(\tau_2)$. 
In the close vicinity of ${\cal C}$ one can introduce the new
set of coordinates $(\tau,s)$,  where $\tau$ runs along ${\cal C}$ and $s$ is the direction which is perpendicular to ${\cal C}$,
as shown on Fig. \ref{fig:C}.
Let ${\bf e}_\tau \equiv (\partial_\tau \tau_1(\tau,s), \partial_\tau \tau_2(\tau,s))$ and ${\bf e}_n \equiv (\partial_s \tau_1(\tau,s), \partial_s \tau_2(\tau,s))$ be the corresponding tangential and transverse vector fields to the curve ${\cal C}$, which in the new
coordinates is just a straight line $s=0$.  According to their definitions we have
\begin{eqnarray}
&&\partial_\tau [f_a(\tau_1(\tau,s))- f_b(\tau_2(\tau,s))] = (f_a'(\tau_1) , - f_b'(\tau_2))^T \cdot {\bf e}_\tau = 0, 
\end{eqnarray}
which means that ${\bf e}_\tau$ should be orthogonal to the vector ${\bf f}_{ab} = (f_a'(\tau_1) , - f_b'(\tau_2))^T$. 
On other hand ${\bf e}_n$ must be parallel to the same vector ${\bf f}_{ab}$.
Therefore, the normalized tangential and normal vectors to the line ${\cal C}$ can be defined by the following relations 
\begin{equation}
\label{transform}
{\bf e}_{\tau}= \frac{(f_b', f_a')}{\sqrt{(f'_a)^2+(f'_b)^2}}, \qquad 
{\bf e}_{n}= \frac{(f_a', -f_b')}{\sqrt{(f'_a)^2+(f'_b)^2}}. 
\end{equation} 
From here it follows that the parametrization  of the coordinate line $s$ in the perpendicular direction ${\bf e}_n$ 
can be written in the following form
\begin{equation}
\left(\begin{array}{c} \tau_1 \\ 
\tau_2 
\end{array}\right) \rightarrow 
\left(\begin{array}{c} \tau_1 \\ 
\tau_2 
\end{array}\right)+ 
\frac{s}{\sqrt{(f'_a)^2+(f'_b)^2}} \left(\begin{array}{c} 
f_a' \\ 
 -f_b'
\end{array}\right) 
\label{perp_shift}
\end{equation}
The difference $f_a(\tau_1)-f_b(\tau_2)$ transforms under Eq. (\ref{perp_shift}) to 
\begin{equation}
f_a\left(\tau_1+s\frac{f_a'}{\sqrt{(f'_a)^2+(f'_b)^2}}\right)-f_b\left(\tau_2-s\frac{f_a'}{\sqrt{(f'_a)^2+(f'_b)^2}}\right)\simeq 
s\sqrt{(f'_a)^2+(f'_b)^2}, 
\label{fa-fb}
\end{equation} 
where we took into account that $f_a(\tau_1)=f_b(\tau_2)$. Now using the exponential parametrization 
\begin{equation}
f'_a=e^{\phi_a}, \, \, \, f'_b=e^{\phi_b}, 
\end{equation}
and substituting Eq. (\ref{fa-fb}) into the integration kernel in Eq. (\ref{eq:S2_reg}), we obtain 
\begin{equation}
S_2[f]=- \frac{N}{32\sqrt{\pi} J} \sum_{a\neq b} g^2 \int\limits_{\sim 1/J}^{+\infty}  \frac{ ds}{s^3} \, \!\!\! 
\int\limits_{\cal C} 
\frac{ d\tau }{ \cosh^{3/2} \left [\phi_a(\tau_1(\tau))- \phi_b(\tau_2(\tau))\right ]}.
\label{S2_full}
\end{equation}
Here we took into consideration that the Jacobian of transformation from $(\tau_1,\tau_2)$ to $(s,\tau)$ variables is unity,
because the basis  $({\bf e}_\tau, {\bf e}_n)$ is orthonormal.  

Considering infinitesimally close transformations in different replicas,  we can expand the denominator in small difference 
$\varphi(\tau)=\phi_a(\tau_1(\tau))- \phi_b(\tau_2(\tau))$, and after performing the integration over $s$, we obtain 
\begin{equation}
S_2[f]\simeq - \frac{ N J }{2^7 \sqrt{2\pi}} \sum_{a\neq b} g^2\int\limits_{\cal C}  
 d\tau  \left(1-\frac{3}{4}\varphi^2\right).
\label{S2_mass_A}
\end{equation}
For the 2nd term in the expression, $\propto \varphi_{ab}^2$, 
the contour ${\cal C}$ can be substituted by the straight line ($\tau_1 = \tau_2 =\tau$)  
if one is interested in the Gaussian order only.
As to the 1st part, the line integral 
\begin{equation}
L_{ab} = \int_{\cal C} dl 
\end{equation}
by itself has a $\varphi^2$--contribution 
which takes into account the deviation of ${\cal C}$ from a straight line. 
As shown below the length reads
\begin{equation}
\label{eq:Lab}
L_{\alpha\beta} =\sqrt{2} \int d\tau \left( 1 +  \frac 18 \varphi_{ab}^2(\tau) + {\cal O}(\varphi_{ab}^3) \right),
\end{equation}
and therefore the final result for the action in the Gaussian order in fluctuations assumes the form
\begin{equation}
S_2[f] = \frac{5 N J }{2^{10} \sqrt{2\pi}}  \sum_{a\neq b} 
g^2_{ab} \int  d\tau \varphi_{ab}^2(\tau) + {\cal O}(\varphi_{ab}^3),
\end{equation}
which is the last expression in Eq.~(\ref{S2_mass}).

Let us now derive Eq.~(\ref{eq:Lab}).  We assume that both phases are small ($\phi_a \ll 1$ 
and $\phi_b \ll 1$) and then parametrize ${\cal C}$ by a variable $\tau$ as
\begin{equation}
\tau_1(\tau) = \tau + x(\tau), \qquad \tau_2(\tau) = \tau + y(\tau),
\end{equation} 
with fluctuations $x,y$ being small in $\phi$'s . 
Their role is to take into account a deviation of the curve from the straight diagonal. From the relation
$f_a(\tau + x) = f_b(\tau + y) $ we then have (up to 2nd order in $\phi$),
\begin{equation}
f_a + e^{\phi_\alpha} x   = f_b + e^{\phi_\beta} y,
\end{equation}
where we took into account that next order terms, e.g $f''_a x^2 \simeq \phi'_a e^{\phi_\alpha} x^2$, 
are of cubic order in $\phi$'s. There are many ways to parametrize the same curve and thus equation above does 
not fix $x$ and $y$ unambiguously. For example one can choose
\begin{equation}
x = -\frac 12 e^{-\phi_a} (f_a - f_b), \qquad
y = \frac 12 e^{-\phi_b} (f_a - f_b).
\end{equation}
With this choice one further needs to evaluate
\begin{equation}
L = \int_{\cal C} dl = \int d\tau \Bigl[(1 + {x}'_\tau)^2  + (1 + {y}'_\tau)^2\Bigr]^{1/2}.
\end{equation}
In the 2nd order in $\phi$'s one finds
\begin{eqnarray}
1 + {x}'_\tau &=& \frac 12 \left[   1 +e^{\phi_b- \phi_a }+ {\phi}'_a (f_a - f_b)  \right], \\
1 + {y}'_\tau &=& \frac 12 \left[   1 +e^{\phi_a - \phi_b }- {\phi}'_b (f_a - f_b)  \right],
\end{eqnarray}
which gives us for the line element
\begin{equation}
dl  =  \sqrt{2} 
\left[ 1 + \frac 38 \varphi_{ab}^2 + \frac 14 \varphi_{ab}' (f_a - f_b)  \right] d\tau.
\end{equation} 
When performing the integral over $d\tau$ we integrate by parts. Taking into account that $f_a' =e^{\phi_a}$ 
we finally arrive at
\begin{equation}
L_{ab} \simeq \sqrt{2} \int d\tau \left( 1 +  \frac 18 \varphi_{ab}^2(\tau) \right),
\end{equation}
as it was claimed above. It is natural that the correction to the geometric length of the curve ${\cal C}$ is positive, 
since it deviates from the straight line. 

\section{Transition to the two-level regime}

In this Appendix we give a qualitative explanation as to why the transition from the reparametrization-dominated to the two-level regime  occurs at a shorter time-scale for the site off-diagonal correlation functions  in comparison to the site-diagonal ones, as observed numerically. 
Consider first the correlation functions $D_2(\tau)$ and $\langle G_{ii}^2(\tau)\rangle_{\mathrm{dis}}$. 
In the two-level regime, with the levels denoted as $|0\rangle$ and $|1\rangle$,  we have 
\begin{equation}
D_2(\tau)=\left\langle \langle 0|\chi_i|1\rangle \langle 1|\chi_j|0\rangle\langle 0|\chi_j|1\rangle \langle 1|\chi_i|0\rangle 
e^{-2E_1\tau}\right\rangle_{\mathrm{dis}}=\left\langle |\langle 0|\chi_i|1\rangle|^2 |\langle 0|\chi_j|1\rangle|^2 e^{-2E_1\tau}\right\rangle_{\mathrm{dis}}
\label{D2_2level}
\end{equation}
Here the energy of the ground state is set to zero hence $E_1$ denotes the energy gap between the ground state and the first excited state.  The numerical data in the two-level regime can be fitted with very high accuracy under the assumptions that  energies $E_1$ and the matrix elements $M_i=\langle 0|\chi_i|1\rangle$ are statistically independent Gaussian distributed quantities. Furthermore, to explain the different time scales for the crossover between the reparametrization dominated and two-level regimes, assume the   matrix elements for the operators at different sites to be statistically independent of each other,  
$\langle M_i M_j \rangle_{\mathrm{dis}}=0$. Then, for the correlation function  $D_2(\tau)$, we obtain 
\begin{equation}
D_2(\tau)=\left(\left\langle |M|^2 \right\rangle_{\mathrm{dis}}\right)^2 \left\langle e^{-2E_1\tau}\right\rangle_{\mathrm{dis}}, 
\label{D2_assumption}
\end{equation}
where $\left\langle |M|^2\right\rangle_{\mathrm{dis}}=\left\langle |M_i|^2 \right\rangle_{\mathrm{dis}}$ independently of $i$.
For $\langle G_{ii}^2(\tau)\rangle_{\mathrm{dis}}$ we obtain under the same assumptions 
\begin{equation}
\langle G_{ii}^2(\tau)\rangle_{\mathrm{dis}}=\left\langle |M|^4 \right\rangle_{\mathrm{dis}} \left\langle e^{-2E_1\tau}\right\rangle_{\mathrm{dis}}.  
\label{G2_assumption}
\end{equation}
Under the assumption of Gaussian distributed matrix elements the results Eqs. (\ref{D2_assumption}) and (\ref{G2_assumption}) differ just by an $N$-independent factor. 

The crossover from the reparametization dominated to the two-level regime occurs at the time scale, where the contributions of higher energy levels become suppressed by the corresponding energy exponents $\sim e^{-E_n\tau}$. As a toy model, consider the contribution of the next exited level, which we denote as $|2\rangle$ with the energy $E_2$. For the correlation function $D_2(\tau)$ we now obtain 
\begin{eqnarray}
\nonumber && 
D_2(\tau)=\left\langle \left(\langle 0|\chi_i |1\rangle \langle 1|\chi_j |0\rangle e^{-E_1\tau}  +\langle 0|\chi_i |2\rangle \langle 2|\chi_j |0\rangle e^{-E_2\tau}\right) \left(\langle 0|\chi_j |1\rangle \langle 1|\chi_i |0\rangle e^{-E_1\tau}  
+\right. \right.\\ 
\nonumber && 
\left.\left.
\langle 0|\chi_j |2\rangle \langle 2|\chi_i |0\rangle e^{-E_2\tau}\right)\right\rangle_{\mathrm{dis}} =\\
\nonumber && 
\left\langle | \langle 0 |\chi_i | 1 \rangle |^2 |\langle 0|\chi_j |1\rangle|^2 e^{-2E_1\tau} +|\langle 0|\chi_i |2\rangle |^2 
|\langle 0|\chi_j |2\rangle|^2 e^{-2E_2\tau}  \right\rangle_{\mathrm{dis}}+  \\ 
&& \left\langle
\left(\langle 0 |\chi_i |1\rangle \langle 1|\chi_j |0\rangle  \langle 0|\chi_j |2\rangle  \langle 2|\chi_i |0\rangle 
+ \langle 0|\chi_j |1 \rangle \langle 1|\chi_i |0\rangle \langle 0|\chi_i |2\rangle \langle 2|\chi_j |0\rangle    \right) 
e^{-(E_1+E_2)\tau} \right\rangle_{\mathrm{dis}}.
\label{D2_3level}
\end{eqnarray}
Eq. (\ref{D2_3level}) is to be contrasted to the 3-level expression for $\langle G_{ii}^2(\tau)\rangle_{\mathrm{dis}}$ 
\begin{eqnarray}
\nonumber && 
\langle G_{ii}^2(\tau)\rangle_{\mathrm{dis}} 
=\left\langle | \langle 0 |\chi_i | 1 \rangle |^4  e^{-2E_1\tau} +|\langle 0|\chi_i |2\rangle |^4 
e^{-2E_2\tau}  \right\rangle_{\mathrm{dis}}+  \\ 
&&  
2 \left\langle
\langle 0 |\chi_i |1\rangle \langle 1|\chi_i |0\rangle  \langle 0|\chi_i |2\rangle  \langle 2|\chi_i |0\rangle 
e^{-(E_1+E_2)\tau} \right\rangle_{\mathrm{dis}}.
\label{G2_3level}
\end{eqnarray}
Now let us make the following assumptions: matrix elements between the ground state and the state $|1\rangle$, and between the ground state and the state $|2\rangle$ are statistically independent. Furthermore, the sign of the product $\langle 0|\chi_j |n \rangle \langle n|\chi_i |0\rangle$, ($i\neq j$) is random, hence the average of such a product over disorder distribution is close to zero. It follows from this assumption, that the terms $\langle 0 |\chi_i |1\rangle \langle 1|\chi_j |0\rangle  \langle 0|\chi_j |2\rangle  \langle 2|\chi_i |0\rangle $ vanish after the average over disorder, in contrast to $\langle 0 |\chi_i |1\rangle \langle 1|\chi_i |0\rangle  \langle 0|\chi_i |2\rangle  \langle 2|\chi_i |0\rangle$, which result in the explicitly positive contribution.   
Eqs. (\ref{D2_3level}), (\ref{G2_3level}) become 
\begin{eqnarray}
 && 
D_2(\tau)\approx  
\left\langle |M_1 |^2 \right\rangle_{\mathrm{dis}}^2  \left\langle e^{-2E_1\tau}\right\rangle_{\mathrm{dis}} + \left\langle |M_2 |^2 \right\rangle_{\mathrm{dis}}^2  \left\langle e^{-2E_2\tau} \right\rangle_{\mathrm{dis}}.
\label{D2_3level-res} \\ 
\nonumber && 
\langle G_{ii}^2(\tau)\rangle_{\mathrm{dis}} 
\approx \left\langle |M_1 |^4 \right\rangle_{\mathrm{dis}} \left\langle e^{-2E_1\tau} \right\rangle_{\mathrm{dis}} +\left\langle |M_2 |^4 \right\rangle_{\mathrm{dis}} \left\langle  e^{-2E_2\tau} \right\rangle_{\mathrm{dis}} +  \\ 
&& 
2 \left\langle |M_1 |^2\right\rangle_{\mathrm{dis}} \left\langle |M_2 |^2 \right\rangle_{\mathrm{dis}} \left\langle e^{-(E_1+E_2)\tau} \right\rangle_{\mathrm{dis}}.
\label{G2_3level-res}
\end{eqnarray}
Here we denoted $M_1=\langle 0 |\chi_i |1\rangle$, $M_2=\langle 0 |\chi_i |2\rangle$. Taking into account $E_1<E_2$, on can see that the subleading term  $\left\langle e^{-(E_1+E_2)\tau} \right\rangle_{\mathrm{dis}}$ in Eq. (\ref{G2_3level-res}) leads to a slower time decay of the site diagonal correlation function $\langle G_{ii}^2(\tau)\rangle_{\mathrm{dis}}$  at relatively short times than the decay of the site off-diagonal correlation function, where the above mentioned term is absent. 

Another, complementary point of view on the spread of the correlation functions in the crossover region between the reparametrization-dominated and two-level regime can be gained by considering the effective many particle density of states. Namely, assuming the discrete spectrum of energies $E_n$ we conclude that the possible energy factors determining the time decay of the site off-diagonal correlation functions can be only the multiples of the energies $E_n$ (in our previous example $n=1, 2$), such as $2E_1, 2E_2, ... $. This is due to the vanishing disorder averages of the matrix elements $\left\langle \langle 0 |\chi_i |n\rangle \langle n|\chi_j |0\rangle \right\rangle_{\mathrm{dis}}$ for $i\neq j$. In contrast, for the site diagonal correlation function, the energy factors are built out of all possible sums of pairs of energies, such as $E_n+E_m$, for any two states $|n\rangle$ and $|m\rangle$. Considering each energy factor as an effective multi-particle energy level, we conclude, that the many-particle energy spectrum contributing to the site-diagonal correlation function is more dense, that is it has lower many-particle level spacing. Now, the deviations from the reparametrization dominated regime should happen at the times which are of the order of inverse many-particle level spacing. According to the considerations above, those times are larger for the site-diagonal correlation function that for the site off-diagonal one. This would explain qualitatively why the spreading of the site off-diagonal curves $\left[D_{2p}/(2p-1)!!\right]^{1/p}$ in Fig. \ref{fig_D2p-N} occurs at earlier times than that for the site-diagonal curves $\left\langle G^p\right\rangle_{\mathrm{dis}}^{1/p}$ in Fig. \ref{fig:Gp}.

\bibliographystyle{JHEP}
\bibliography{SYK}


\end{document}